\documentclass{nature}
\usepackage{float}
\usepackage{graphicx}
\usepackage{pdflscape}
\usepackage{hyperref}
\usepackage{pgf,pgffor}
\usepackage{textcomp}
\usepackage{threeparttable} 
\usepackage{xcolor}
\usepackage{ulem}
\usepackage{lineno}
\usepackage{caption}
\usepackage[caption = false]{subfig}
\usepackage{pifont}
\usepackage{longtable}
\usepackage{amssymb}
\usepackage{cleveref}
\definecolor{orcidlogocol}{HTML}{A6CE39}
\usepackage[symbol]{footmisc}

\makeatletter

\let\saved@includegraphics\includegraphics
\AtBeginDocument{\let\includegraphics\saved@includegraphics}
\renewenvironment*{figure}{\@float{figure}}{\end@float}
\makeatother

\newcommand{\EQ}[1] {Eq.~(\ref{#1})}
\newcommand{\FIG}[1] {Figure~\ref{#1}}

\newcommand{\EXTFIG}[1] {Extended Data Figure~\ref{#1}}

\newcommand{\AFF}[1]{$^{\foreach\d[count=\ni]in{#1}{\ifnum\ni=1\ref{\d}\else,\ref{\d}\fi}}$}

\newcommand{\ion}[2]{[#1\,\textsc{#2}]}

\def\DM{\mathrm{DM}}
\def\RM{\mathrm{RM}}
\def\DMm{\DM_{\rm MW}}

\def\DMi{\DM_{\rm IGM}}
\def\DMh{\DM_{\rm host}}

\def \arcsec {''}

\def\cmpc{\mathrm{cm}^{-3}\, \mathrm{pc}}
\def\radm{\mathrm{rad}\,\mathrm{m}^{-2}}

\definecolor{dkblue}{RGB}{54, 86, 169}
\newcommand{\REV}[1] {#1}

\title{A fast radio burst source at a complex magnetised site in a barred 
galaxy}
\author{H.~Xu\AFF{aff:KIAA,aff:NAOC,aff:PKU},J.~R.~Niu\AFF{aff:NAOC,aff:UCAS}, P.~Chen\AFF{aff:KIAA, aff:PKU, aff:WIS},
K.~J.~Lee\AFF{aff:KIAA, aff:NAOC}\thanks{E-mail: kjlee@pku.edu.cn,\href{https://orcid.org/0000-0002-1435-0883}{\textcolor{orcidlogocol}{} \hspace{2mm} orcid.org/0000-0002-1435-0883}}, 
W.~W.~Zhu\AFF{aff:NAOC}\thanks{Email: zhuww@nao.cas.cn,\href{https://orcid.org/0000-0001-5105-4058}{\textcolor{orcidlogocol}{} \hspace{2mm} orcid.org/0000-0001-5105-4058}},
S.~Dong\AFF{aff:KIAA}\thanks{Email: dongsubo@pku.edu.cn,\href{https://orcid.org/0000-0002-1027-0990}{\textcolor{orcidlogocol}{} \hspace{2mm} orcid.org/0000-0002-1027-0990 }}, 
B.~Zhang\AFF{aff:NCfA, aff:UNLV}\thanks{Email: bing.zhang@unlv.edu,\href{https://orcid.org/0000-0002-9725-2524}{\textcolor{orcidlogocol}{} \hspace{2mm} orcid.org/0000-0002-9725-2524 }},
J.~C.~Jiang\AFF{aff:KIAA, aff:NAOC,aff:PKU},  B.~J.~Wang\AFF{aff:KIAA, aff:NAOC,aff:PKU},  J.~W.~Xu\AFF{aff:KIAA, aff:NAOC,aff:PKU}, C.~F.~Zhang\AFF{aff:KIAA, aff:NAOC,aff:PKU}, H.~Fu\AFF{aff:IOWA}, A.~V.~Filippenko\AFF{aff:UCB}, E.~W.~Peng\AFF{aff:PKU,aff:KIAA}, D.~J.~Zhou\AFF{aff:NAOC,aff:UCAS}, Y.~K.~Zhang\AFF{aff:NAOC,aff:UCAS},
P.~Wang\AFF{aff:NAOC},
Y.~Feng\AFF{aff:NAOC,aff:ZJL}, Y.~Li\AFF{aff:pmo}, T.~G.~Brink\AFF{aff:UCB}, D.~Z.~Li\AFF{aff:calt}, W.~Lu\AFF{aff:PRIN}, Y.~P.~Yang\AFF{aff:swifa},
R.~N.~Caballero\AFF{aff:KIAA}, C.~Cai\AFF{aff:IHEP}, M.~Z.~Chen\AFF{aff:xao},
Z.~G.~Dai\AFF{aff:USTC}, S.~G.~Djorgovski\AFF{aff:Caltech}, A.~Esamdin\AFF{aff:xao}, H.~Q.~Gan\AFF{aff:NAOC}, P.~Guhathakurta\AFF{aff:UCSC}, J.~L.~Han\AFF{aff:NAOC},
L.~F.~Hao\AFF{aff:YNAO}, Y.~X.~Huang\AFF{aff:YNAO},
P.~Jiang\AFF{aff:NAOC}, C.~K.~Li\AFF{aff:IHEP}, D.~Li\AFF{aff:NAOC,aff:gzn},H.~Li\AFF{aff:NAOC}, X.~Q.~Li\AFF{aff:IHEP}, Z.~X.~Li\AFF{aff:YNAO}, Z.~Y.~Liu\AFF{aff:xao}, R.~Luo\AFF{aff:csiro},  Y.~P.~Men\AFF{aff:mpifr}, C.~H.~Niu\AFF{aff:NAOC}, W.~X.~Peng\AFF{aff:IHEP},  L.~Qian\AFF{aff:NAOC}, L.~M.~Song\AFF{aff:IHEP}, D.~Stern\AFF{aff:jpl}, A.~Stockton\AFF{aff:HAWAII}, J.~H.~Sun\AFF{aff:NAOC}, F.~Y.~Wang\AFF{aff:nju},  M.~Wang\AFF{aff:YNAO}, N.~Wang\AFF{aff:xao}, W.~Y.~Wang\AFF{aff:PKU}, X.~F.~Wu\AFF{aff:pmo}, S.~Xiao\AFF{aff:IHEP}, S.~L.~Xiong\AFF{aff:IHEP}, Y.~H.~Xu\AFF{aff:YNAO}, R.~X.~Xu\AFF{aff:PKU,aff:KIAA,aff:pklab}, J.~Yang\AFF{aff:nju}, X.~Yang\AFF{aff:pmo}, R.~Yao\AFF{aff:NAOC}, Q.~B.~Yi\AFF{aff:IHEP}, Y.~L.~Yue\AFF{aff:NAOC}, D.~J.~Yu\AFF{aff:NAOC}, W.~F.~Yu\AFF{aff:shao}, J.~P.~Yuan\AFF{aff:xao}, B.~B.~Zhang\AFF{aff:nju,aff:klma}, S.~B.~Zhang\AFF{aff:pmo}, S.~N.~Zhang\AFF{aff:IHEP}, Y.~Zhao\AFF{aff:IHEP},  W.~K.~Zheng\AFF{aff:UCB}, Y.~Zhu\AFF{aff:NAOC}, J.~H.~Zou\AFF{aff:hbu,aff:nju}}

\begin{document}
\maketitle

\begin{affiliations}
\item Kavli Institute for Astronomy and Astrophysics, Peking University, Beijing 100871, P.~R.~China \label{aff:KIAA}
\item National Astronomical Observatories, Chinese Academy of Sciences, Beijing 100101, P.R.~China\label{aff:NAOC}
\item Department of Astronomy, Peking University, Beijing 100871, P.~R.~China\label{aff:PKU}
\item University of Chinese Academy of Sciences, Chinese Academy of Sciences, Beijing 100049, P.~R.~China\label{aff:UCAS}
\item Department of Particle Physics and Astrophysics, Weizmann Institute of Science, Rehovot 76100, Israel\label{aff:WIS}
\item Nevada Center for Astrophysics, University of Nevada, Las Vegas, NV 89154, USA\label{aff:NCfA}
\item Department of Physics and Astronomy, University of Nevada, Las Vegas, NV 89154, USA\label{aff:UNLV}
\item Department of Physics \& Astronomy, University of Iowa, Iowa City, IA 52242, USA\label{aff:IOWA}
\item Department of Astronomy, University of California, Berkeley, CA 94720-3411, USA\label{aff:UCB}
\item {Zhejiang Lab, Hangzhou, Zhejiang 311121, People's Republic of China} \label{aff:ZJL}
\item Purple Mountain Observatory, Chinese Academy of Sciences, Nanjing 210008, P~.R~. China\label{aff:pmo}
\item TAPIR, Walter Burke Institute for Theoretical Physics, Mail Code 350-17, Caltech, Pasadena, CA 91125, USA\label{aff:calt}
\item Department of Astrophysical Sciences, Princeton University, Princeton, NJ 08544, USA\label{aff:PRIN}
\item South-Western Institute For Astronomy Research, Yunnan University, Yunnan 650504, P.~R.~China\label{aff:swifa}
\item Key laboratory of Particle Astrophysics, Institute of High Energy Physics, Chinese Academy of Sciences, Beijing 100049, P.R.~China\label{aff:IHEP}
\item Xinjiang Astronomical Observatory, Chinese Academy of Sciences, Urumqi 830011, P.~R.~China\label{aff:xao}
\item University of Science and Technology of China, Anhui 230026, P.~R.~China.\label{aff:USTC}
\item Division of Physics, Mathematics, and Astronomy, California Institute of Technology, Pasadena, CA 91125, USA.\label{aff:Caltech}
\item UCO/Lick Observatory, Department of Astronomy \& Astrophysics, University of California Santa Cruz, 1156 High Street, Santa Cruz, CA 95064, USA\label{aff:UCSC}
\item Yunnan Observatories, Chinese Academy of Sciences, Kunming 650216, P.~R.~China\label{aff:YNAO}
\item Guizhou Normal University, Guiyang 550001, P.~R.~China \label{aff:gzn}
\item CSIRO Space and Astronomy, Epping, NSW 1710, Australia\label{aff:csiro}
\item Max-Planck institut für Radioastronomie, Auf Dem Hügel, Bonn, 53121, Germany\label{aff:mpifr}
\item Jet Propulsion Laboratory, California Institute of Technology, 4800 Oak Grove Drive, Pasadena, CA 91109, USA\label{aff:jpl}
\item Institute for Astronomy, University of Hawaii, Honolulu, HI 96822, USA\label{aff:HAWAII}
\item School of Astronomy and Space Science, Nanjing
University, Nanjing 210093, P.~R.~China\label{aff:nju}
\item State Key Laboratory of Nuclear Physics and Technology, School of Physics, Peking University, Beijing 100871, P.~R.~China \label{aff:pklab}
\item Key Laboratory of Modern Astronomy and Astrophysics (Nanjing University), Ministry of Education, P.~R.~China\label{aff:klma}
\item College of Physics, Hebei Normal University, Shijiazhuang 050024, P.~R.~China\label{aff:hbu}
\item Shanghai Astronomical Observatory, Chinese Academy of Sciences, Shanaghai 200030, P.~R.~China\label{aff:shao}
\end{affiliations}

\begin{abstract}
Fast radio bursts (FRBs) are highly dispersed millisecond-duration radio bursts\cite{Petroff19AAR,Cordes19ARAA,Zhang20Nature}.
Recent observations of a Galactic FRB\cite{CHIME200428,Bocheneck20,Li21,Ridnaia21,Mereghetti20} \REV{suggest} that at least some FRBs originate from magnetars, but the origin of cosmological FRBs
is still not settled.
Here we report the detection of 1863 
bursts in 82 hr over 54 days from the repeating source FRB~20201124A\cite{CHIMEFRB20201124A}. 
These observations show
irregular short-time variation of the Faraday rotation measure (RM), which probes the density-weighted line-of-sight magnetic field strength, of individual bursts during the first 36\,days, followed by a constant RM.
We detected circular polarisation in more than half of the burst sample, including one burst \REV{reaching a high fractional circular polarisation of 75\%}.
Oscillations in 
\REV{fractional} linear and circular polarisations as well as polarisation angle as a function of wavelength were detected. 
All of these features provide evidence for a complicated, dynamically evolving, magnetised immediate environment 
\REV{within} about an astronomical unit (au; Earth-Sun distance)  \REV{of} the source.
Our optical observations of its Milky-Way-sized, metal-rich host galaxy\cite{Fong21, Ravi21, Piro21}
reveal a barred spiral, 
with the FRB source residing in a low stellar density, interarm region at an intermediate galactocentric distance. 
This environment is inconsistent with a young magnetar engine formed during an extreme explosion of a massive star that resulted in a long gamma-ray burst or superluminous supernova. 
\end{abstract}

Triggered by observations of the Canadian Hydrogen Intensity Mapping Experiment\cite{CHIMEFRB20201124A}, we used the Five-hundred-meter Aperture Spherical radio Telescope (FAST)\cite{Jiang19SCPMA} to monitor FRB~20201124A from 2021 April 1 to June 11 (UT) with 91\,hr total observing time. The 19-beam receiver was used to cover the frequency range 1.0--1.5 GHz. Our detection threshold was a signal-to-noise ratio $S/N>7$, and 1103 bright bursts reached $S/N>30$ 
among a total of 1863 detected bursts. The burst flux is 0.005--11.5\,Jy, and the inferred isotropic luminosity after integrating signal bandwidth spans from $5\times10^{37}$\,erg\,s$^{-1}$ to $3\times10^{40}$\,erg\,s$^{-1}$. The daily burst energy distribution shows {no secular trend} (Methods), while the burst-to-burst fluctuation exceeds two orders of magnitude. 

\begin{figure}
\centering
\includegraphics[width=1.0\textwidth]{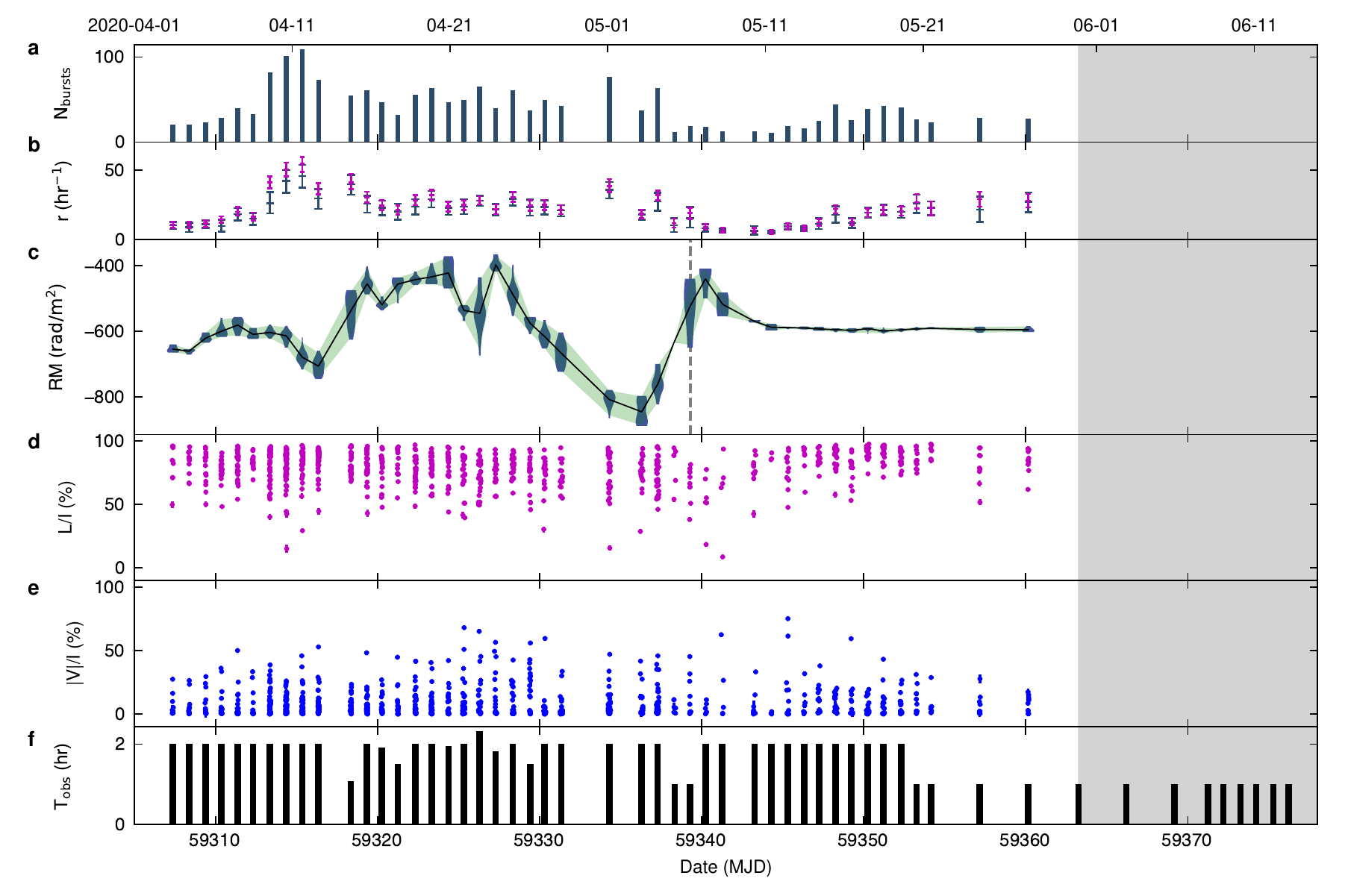}
\caption{\textbf{Temporal variations of the physical parameters of FRB~20201124A.} {\bf (a)} Daily number of bursts detected. {\bf (b)} Weibull event rate (blue) and Poisson event rate (magenta). {\bf (c)} Daily RM, where the violin symbol indicates the distribution function, the green shaded strips indicate the 95\% upper and lower bounds, and the solid black curve is the median. Vertical dashed line indicates MJD~59339, where the largest RM fluctuation is noted. {\bf (d)} and {\bf (e)} Fractional linear and circular polarisations measured for each individual burst. {\bf (f)} Observation length \REV{on} each day. The grey shaded region on the right side of the plot shows the epoch when no burst was detected. 
\label{fig:fig1}
}
\end{figure}

The daily event rate varies slowly (\FIG{fig:fig1}), with minimal and maximal values of $6\pm1$ and $46\pm8$\,hr$^{-1}$, respectively. \REV{Throughout paper, the error bars are for the 68\% confidence level unless otherwise specified.} The waiting time follows a bimodal distribution with timescales peaking at 39\,ms and 135.2\,s (Methods). {Similar bimodality in waiting time had also been detected in the repeating FRB~20121102A\cite{LiDi2021},
which may indicate a common mechanism.} The high event rate makes FRB~20201124A among the most active known FRBs. {We witnessed the quenching of the burst activity on a timescale $<74$\,hr, when the source stopped emitting any bursts above the flux limit of
4.3\,mJy at the fiducial burst width of 5\,ms on 2021 May 29.} We continued to observe the source over the next 16\,days and did not detect \REV{a single burst} during the 9\,hr of observations (\FIG{fig:fig1}). {Counterintuitively, the burst rate {did not show} any sign of a monotonic decrease, but a \REV{slow} increase from $6\pm1$\,hr$^{-1}$ to $27_{-8}^{+7}$\,hr$^{-1}$ during the last 20\,days before the quenching.}

\begin{figure} 
\centering
\includegraphics[width=\textwidth]{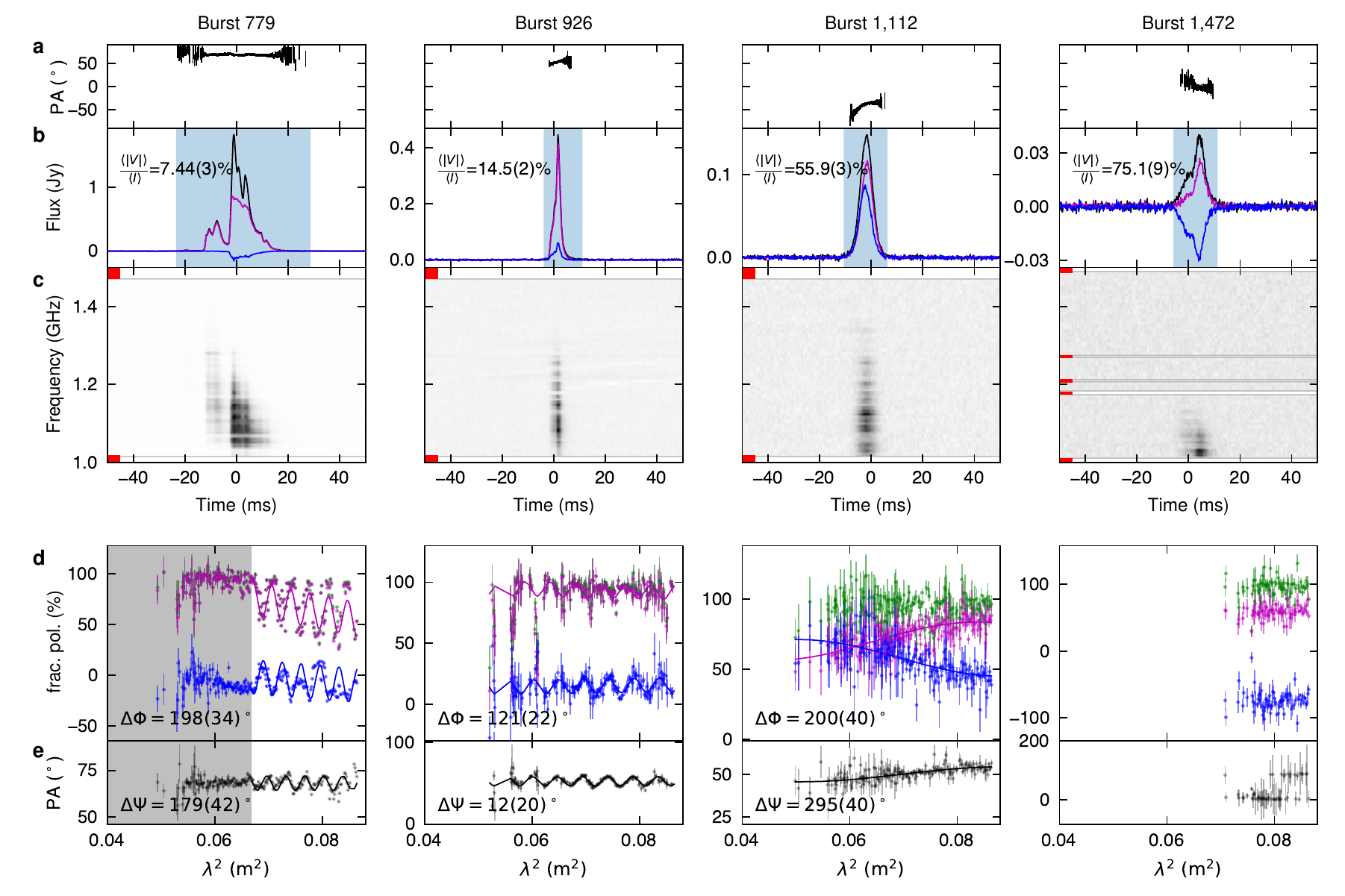}
\caption{\textbf{Polarisation profiles, dynamic spectra, and frequency-dependent polarisation of selected bursts.} {\bf (a)} PA curve. {\bf (b)} Polarisation profile, where total intensity, linear, and circular polarisations normalised to the off-pulse noise of total intensity are in black, red, and blue curves, respectively. {\bf (c)} Dynamic spectra of total intensity. The horizontal white strips and red markers represent frequency channels that have been removed owing to either radio-frequency interference (RFI) or band edges. {\bf (d)} Fractional polarisation as a function of the square of wavelength, where green, magenta, and blue dots and error bars are respectively for total, linear, and circular polarisations. The solid curves of the corresponding colour are the model fitting excluding data in the grey region (see Methods). The oscillation phase difference ($\Delta \Phi$) between the linear and circular polarisation oscillations is denoted. {\bf (e)} Linear polarisation angle as a function of the square of wavelength. The oscillation phase difference ($\Delta \Psi$, between the linear polarisation and polarisation angle is denoted. \REV{All error bars in the figure are at the 95\%-confidence level.}
\label{fig:fig2}
}
\end{figure}

FRB~20201124A bursts show diverse polarisation properties, including
nearly constant polarisation angle (PA) across the phase\cite{HilmarssonFRB20201124A,Michilli18Nat}, significant PA swings\cite{Luo20Nat}, and a high amount of circular polarisation (\FIG{fig:fig2}). 
In our burst sample, 50\% of the bright bursts ($S/N>30$) had circular polarisation higher than 3.3\%, while the maximal circular polarisation reaches 75\%, higher than the 47\% reported previously\cite{Kumar2021}. This is in contrast to most FRBs\cite{Petroff19AAR} or radio-emitting magnetars\cite{Kramer2007} showing little circular polarisation. 

For some bursts with moderate circular polarisation, the frequency spectra of both circular and linear polarisations show oscillating features (e.g., bursts 779 and 926 in \FIG{fig:fig2}), which indicate Faraday conversion (i.e., generalised Faraday rotation) or polarisation-dependent absorption. The oscillation phases of the linear and circular polarisations are approximately offset by $180^\circ$. We also detected highly circularly polarised bursts without such  quasiperiodic structures (burst 1472 in \FIG{fig:fig2}). This suggests that there is likely an alternative mechanism for producing circular polarisation in addition to the polarisation oscillations.
Since the synchrotron maser model invoking relativistic shocks does not predict circular polarisation, our results 
\REV{support} the magnetospheric origin of FRB emission\cite{Kumar17MN,YangZhang18ApJ,Luo20Nat,Zhang20Nature}.

The RM shows a stochastic temporal variation between $-889.5^{+0.7}_{-0.7}$ and $-365.1^{+2.9}_{-1.4}$\,rad\,m$^{^-2}$, on a timescale of 10\,days (\FIG{fig:fig1}) and the largest burst-to-burst RM variation {per session} with a root-mean-square (RMS) value of 77.2\,rad\,m$^{-2}$.
Within a single burst, the profile evolution induced apparent RM variation is $\sim 15.6$\,rad\,m$^{-2}$. The detected  stochastic RM variations on 10-day timescale are more than 40 times larger, suggesting that stochastic RM variations do not result from the profile evolution. Compared with the case of FRB~20121102A\cite{Michilli18Nat,Hilmarsson2021}, the fractional amplitude of RM variation in FRB~20201124A is larger, while the absolute amplitude is smaller.
The RM variation quickly stopped $\sim 20$\,days before the quenching of the radio bursts, {with the 95\% confidence level upper limit of $\Delta {\rm RM}\le9.1$ \,${\rm rad\,m^{^-2}}$, a factor of 50 smaller than the amplitude of stochastic RM variations detected earlier. }

{The significant RM variation on a 10-day timescale could be caused by a change of either the magnetic field configuration or the density profile along the line of sight close to the source region. 
The sub-au size of the Faraday screen is estimated as $\sim 0.6 \ {\rm au} (\tau / 10 {\rm d}) (v / 100 \ {\rm km~s^{-1}})$,
where $\tau$ is the timescale of RM variation and $v$ is the relative transverse velocity between the screen and the FRB source.} \REV{Similar to the FRB~20201124A, the Galactic binary pulsar system PSR~B1259$-$63 shows irregular RM evolution\cite{Johnston05}. Thus, the lack of periodicity in RM variation may not rule out the binary scenario, and is probably related to irregular mass ejection from the companion star.}
The cessation of RM variation suggests that the line of sight is less contaminated by the varying component of the medium density. 
If the central engine is an isolated young magnetar, the RM is predicted to show a secular monotonic decline with time\cite{Piro18,Metzger19MN}. 

{The oscillations of the linear and circular polarisation \REV{fractions} as well as the linear polarisation angle as functions of radio wavelength probe the magnetic field and the relativistic plasma close to the FRB central engine. 
They are likely a consequence of the polarisation-dependent absorption or conversion, which requires 
$B\ge 3(\gamma/10)^{-2}$\,Gauss, where $\gamma$ is the kinetic Lorentz factor of the plasma (Methods).
Owing to the burst-to-burst variation of polarisation oscillation phenomena, the distance scale associated with the oscillation in polarisation is  $0.1\,{\rm au}\, (\tau/{\rm min})$ assuming that the bursts propagate at the speed of light.
The polarisation oscillation suggests that 
the vicinity of the FRB source is occupied by a variable 
Gauss-level magnetic field together with both a cold and a relativistic plasma (Methods). 
}

\begin{figure} 
\centering
\includegraphics[width=0.5\textwidth]{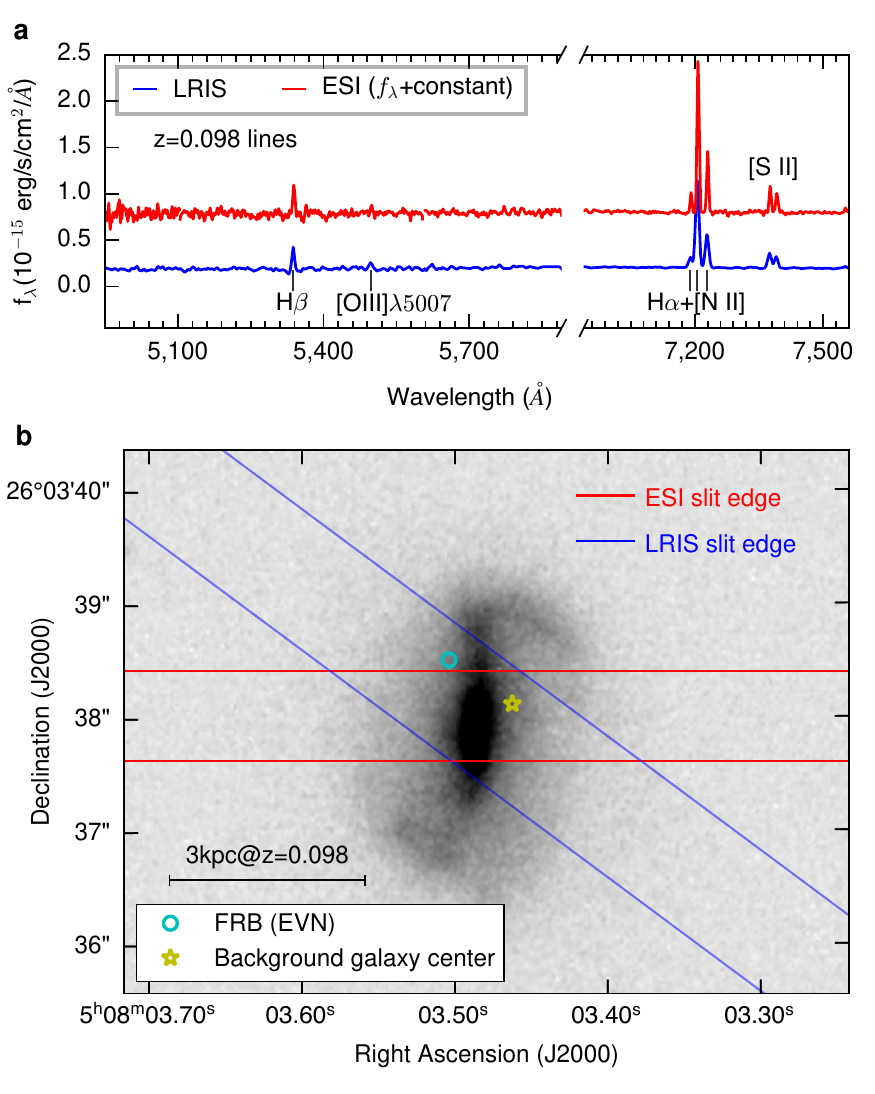}
\caption{\textbf{Host-galaxy properties at optical and near-infrared wavelengths.} {\bf (a)} Emission lines from the $z=0.098$
host galaxy in the LRIS (blue) and ESI (red) spectra.
{\bf (b)} The $K^\prime$-band AO image of the barred-spiral host galaxy, with the indicated position of the FRB\cite{FRB20201124AEVN} shown as a cyan circle, the centroid of a $z=0.553$ background galaxy (Methods) marked by a yellow star, and the LRIS and ESI slit edges in blue and red solid lines, respectively.
\label{fig:fig3}
}
\end{figure}

We performed optical and near-infrared
observations of the host galaxy SDSS 
J050803.48+260338.0\cite{Fong21,Ravi21,Piro21} using 
the 10\,m Keck telescopes, including high- and low-dispersion spectra with the 
Echellette Spectrograph Imager (ESI) and the Low Resolution Imaging Spectrometer 
 (LRIS), respectively, and $K^\prime$-band images with the NIRC2 camera using the laser guide-star adaptive-optics (AO) system.
In the optical spectra, we detected multiple 
emission lines (\FIG{fig:fig3}(a)) and derived
$z=0.09795\pm0.00003$, which corresponds to a luminosity distance of 
$453.3\pm0.1$\,Mpc 
adopting the standard {\it Planck} cosmological model\cite{Planck16A&A}. Similar to the hosts of several other repeaters (e.g., FRB~20121102A\cite{Chatterjee2017}, 
FRB~20180916B\cite{Marcote2020}, FRB~20180301A\cite{Bhandari2021}), this host is 
in the star-forming branch of the Baldwin-Phillips-Terlevich
diagram\cite{BPT1981}  (Methods). Our adaptive-optics (AO) image 
(\FIG{fig:fig3}(b)), with a full width at half-maximum resolution of $0.12''$, shows that the host is a barred galaxy with
spiral features, and the FRB's apparent location is in the disc but offset from 
the bar and spiral arms.

The galaxy's stellar mass\cite{Fong21,Ravi21}, 
$M_*\approx3\times10^{10}\,M_{\odot}$, is about half that of the Milky Way (MW);
in contrast, its 
star-formation rate (${\rm SFR}=3.4\pm0.3\,M_{\odot}\,{\rm yr^{-1}}$) is about 
twice that of the MW, and its metallicity ($12+\log{\rm(O/H)} = 
9.07^{+0.03}_{-0.04}$) is approximately twice the solar abundance (Methods). 
The projected offset of the FRB 
location from the galaxy centre and the specific SFR appear to be typical 
compared with known FRB hosts (Methods), and its metallicity is higher than that of any 
FRB host reported previously (most were for non-repeating FRBs)\cite{Bhandari20ApJL,Li2020}. 
One active repeater, FRB~20121102A, is hosted by a metal-poor dwarf galaxy with a high specific SFR\cite{Chatterjee2017,Tendulkar17ApJ}, which is similar to the hosts of long gamma-ray bursts or hydrogen-poor superluminous supernovae. This has motivated a hypothesised connection between active repeating FRBs and young, millisecond magnetars\cite{Metzger17ApJ}. In contrast, the host of FRB~20201124A is more metal-rich and massive than almost all known hosts of long gamma-ray bursts or hydrogen-poor superluminous supernovae\cite{Li2020}. It has been speculated that FRB~20201124A may reside in a star-forming region in the host galaxy\cite{Fong21,Ravi21,Piro21}. However, the interarm location of the source revealed by our image does not support such a possibility, so a young magnetar engine born from an extreme explosion 
is disfavoured. Nonetheless, a regular magnetar similar to those in the MW is still possible, \REV{although the high burst rate, not seen in the Galactic magnetars, requires unusual intrinsic or environmental conditions.}

\clearpage

\bibliographystyle{naturemag}
\clearpage

\noindent \FIG{fig:fig1}:\textbf{Temporal variations of the physical parameters of FRB~20201124A.} {\bf (a)} Daily number of bursts detected. {\bf (b)} Weibull event rate (blue) and Poisson event rate (magenta).
{\bf (c)} Daily RM, where the violin symbol indicates the distribution function, the green shaded strips indicate the 95\% upper and lower bounds, and the solid black curve is the median. Vertical dashed line indicates MJD~59339, where the largest RM fluctuation is noted. {\bf (d)} and {\bf (e)} Fractional linear and circular polarisations measured for each individual burst. {\bf (f)} Observation length of each day. The grey shaded region on the right side of the plot shows the epoch when no burst was detected. 

\noindent \FIG{fig:fig2}:\textbf{Polarisation profiles, dynamic spectra, and frequency-dependent polarisation of selected bursts.} {\bf (a)} PA curve. {\bf (b)} Polarisation profile, where total intensity, linear, and circular polarisations normalised to the off-pulse noise of total intensity are in black, red, and blue curves, respectively. {\bf (c)} Dynamic spectra of total intensity. The horizontal white strips and red markers represent frequency channels that have been removed owing to either radio-frequency interference (RFI) or band edges. {\bf (d)} Fractional polarisation as a function of the square of wavelength, where green, magenta, and blue dots and error bars are respectively for total, linear, and circular polarisations. The solid curves of the corresponding colour are the model fitting excluding data in the grey region (see Methods). The oscillation phase difference ($\Delta \Phi$) between the linear and circular polarisation oscillations is denoted. {\bf (e)} Linear polarisation angle as a function of the square of wavelength. The oscillation phase difference ($\Delta \Psi$, between the linear polarisation and polarisation angle is denoted. \REV{All error bars in the figure are at the 95\%-confidence level.}

\noindent \FIG{fig:fig3}:\textbf{Host-galaxy properties at optical and near-infrared wavelengths.} {\bf (a)} Emission lines from the $z=0.098$
host galaxy in the LRIS (blue) and ESI (red) spectra.
{\bf (b)} The $K^\prime$-band AO image of the barred-spiral host galaxy, with the indicated position of the FRB\cite{FRB20201124AEVN} shown as a cyan circle, the centroid of a $z=0.553$ background galaxy (Methods) marked by a yellow star, and the LRIS and ESI slit edges in blue and red solid lines, respectively.
\clearpage

\clearpage

\clearpage

\clearpage
\section*{Methods}

\subsubsection*{Radio observations and burst detection}
We carried out our observations using the 19-beam receiver of FAST on 2021 April 1. The 19-beam receiver spans from 1.0\,GHz to 1.5\,GHz with a system temperature of 20--25\,K \cite{Jiang20RAA}. {From 2021 April 1 to April 2, we  performed a grid of 9 observations using all 19 beams around the position ($\alpha =05^{\rm h}08^{\rm m}$, $\delta =+26^\circ11^\prime$) reported by the CHIME/FRB team\cite{CHIMEAtel} and detected multiple bursts in 2 to 4 beams simultaneously. We then used the differential intensity in each beam to compute a refined location: $\alpha = 05^{\rm h}08^{\rm m}03.50^{\rm s}$, $\delta = +26^\circ03^\prime37.80^{\prime\prime}$ \cite{XuAtel2021}. From May 15 onward, observations were carried out by pointing the FAST central beam at the much more precise European Very Long Baseline Interferometry Network (EVN) position\cite{FRB20201124AEVN} ($\alpha = 05^{\rm h}08^{\rm m}03.507^{\rm s}$, $\delta = +26^\circ03^\prime38.50^{\prime\prime}$).  The epochs and durations of all observations are shown in \FIG{fig:fig1}. The data of April 1 and 2 were used only for localisation purposes\cite{XuAtel2021}; they are excluded in other analyses in this paper, as the beam centre was misaligned with the source position.} The data were recorded with a frequency resolution of 122.07\,kHz and a temporal resolution of 49.152\,$\mu$s or 196.608\,$\mu$s. The FAST receiver uses the dual-polarisation linear polarisation feed\cite{Jiang20RAA}, with which the 4-channel Stokes intensity was measured. Before and after each observation session, we recorded a 1~minute noise diode signal for the polarisation calibration. 

We used the software \textsc{transientx} {(\url{https://github.com/ypmen/TransientX})} to perform the off-line burst searches. The data were dedispersed in the range of 380--440\,$\cmpc$ with a step of 0.1\,$\cmpc$ and the burst width was searched with a boxcar filter, of which the filter width ranges from 0.1\,ms to  100\,ms. After candidate plots were formed, we visually inspected all 3364 candidates with $S/N \ge 7$\cite{ZXM21}. A total of 1863 bursts were detected in our observations; the detected number of bursts for each observation session is plotted in \FIG{fig:fig1}. We also verified the search results using the software \textsc{BEAR}\cite{Men19MN}. No difference can be found for bursts with $S/N \ge 7$.

\subsubsection*{Event-rate evolution and the quenching}

We adopted the Weibull distribution\cite{Oppermann18MN} to describe the probability density of time intervals between bursts. 
The Weibull  distribution of time interval $\delta$ is
\begin{equation}
		W(\delta|k,r) = k\delta^{-1}[\delta\,r \Gamma(1+1/k)]^ke^{-[\delta\,r 
		\Gamma(1+1/k)]^k},
\end{equation}
where the function $\Gamma(x)\equiv\int_0^\infty t^{x-1}\mathrm{e}^{-t}\mathrm{d}t$; $r$ and $k$ are the event rate and shape parameter, respectively.
The statistical inferences  were carried out using the Bayesian method\cite{Oppermann18MN} implemented with the software package \textsc{multinest}\cite{Feroz09MN}.
The inferred daily event rate and shape parameter are shown in \FIG{fig:fig1} and \EXTFIG{fig:radiosup1}.  The inferred average event rate and shape parameter are $r=21\pm2\,\mathrm{hr^{-1}}$ and $k=0.60\pm0.02$ for a 95\% confidence level, where $k<1$ indicates that the bursts tend to cluster together. {The Poisson rate, implicitly assuming no temporal correlation, is computed by fixing $k=1$. As shown in \FIG{fig:fig1}, the two rates are consistent with each other within the 68\% confidence-level error. {The event rate 
is not constant as the reduced $\chi^2$, assuming constant rate, for all 45 measurements are 6.3 and 12.0 (the corresponding $P$-values are $10^{-35}$ and $10^{-83}$) for the Weibull and Poission cases, respectively.}} On May 29 (MJD~59363), the FRB source was quenched. No more bursts were detected with $S/N \ge 7$ thereafter in 20\,days with a total of 9\,hr observations. The corresponding 95\% confidence level upper limit of the event rate is $r\le 0.3$\,hr$^{-1}$. 

\subsubsection*{Flux, fluence, and energy of bursts}\label{sec:burstprop}

We estimated the flux densities ($S$) through the radiometer equation
with a typical system temperature of 20\,K and telescope gain $G \approx 16$\,K\,Jy$^{-1}$ for FAST\cite{Jiang20RAA}. We calculate the flux density at a frequency resolution of 7.8125\,MHz and a time resolution of 196.608\,$\mu$s. The dominant uncertainty ($\sim 20\%$) in flux-density estimation comes from the temporal variation of system temperature\cite{Jiang20RAA}. The average flux density is derived from the Gaussian fitting method\cite{CHIME20Nat}. Since the average flux will depend on the definition of ``signal bandwidth'', we choose the 3$\sigma$ width from the Gaussian fitting as our signal bandwidth. The average burst fluence ($F$) is computed by integrating the average burst flux with respect to time, and the equivalent width $W_{\mathrm{eq}}$ is computed by dividing the fluence by the burst peak flux. The measured distributions $F$ and $W_{\mathrm{eq}}$ are shown in \EXTFIG{fig:flux}. The average and the RMS deviation of $W_{\mathrm{eq}}$ are 7.6\,ms and 3.3\,ms, respectively, while the average $F$ and its RMS fluctuation are 0.5\,Jy\,ms and 1.0\,Jy\,ms.

The sample completeness was determined with the following method. We simulated 10,000 mock bursts with Gaussian profile and bandpass matching the detected distributions. We then randomly injected the mock bursts into the original FAST data when no FRB was detected. The mock burst injected data are then fed to our burst-searching pipeline to compute the detection rate. The procedure shows that the fluence threshold achieving the 95\% detection probability with $S/N \ge 7$ is 53\,mJy\,ms.

The isotropic burst energy $E$ was calculated by integrating over the $4\pi$ solid angle and the signal bandwidth, $E=4\pi D_{L}^2(1+z)^{-1}\int F\,d{\rm BW}$,
where $D_{L}=453.3\pm0.1$\,Mpc is the luminosity distance\cite{Planck16A&A}, and $F$ is the fluence. {The energy inference is little affected by the choice of BW, as it is integrated. We evaluate the systematics by comparing the energy measured from integrating 2$\sigma$ and 3$\sigma$ Gaussian-fitting BW. The difference ($\Delta E/E\le $10\%) is smaller than the 20\% uncertainty in the system temperature.}

The histogram of burst energies and the cumulative distribution function (CDF) of the burst energy are shown in \EXTFIG{fig:flux}. {We note that the CDF of burst energy is better fitted by a broken power law rather than by a single power law; i.e., we use
\begin{equation}
N(\ge E)\propto \left\{\begin{array}{c}
    E^{-\gamma_1}\, {\rm for}\,E< E_0\,, \\
    E^{-\gamma_2}\, {\rm for}\, E\ge E_0\,,
\end{array}    \right.
\end{equation}
where $\gamma_1$ and $\gamma_2$ are the power-law indices, and $E_0$ is the turning-over energy.
The inferred parameters for the broken power law model are $\gamma_1=0.36\pm0.02\pm0.02$, $\gamma_2=1.5\pm0.1\pm0.1$, and $E_0=1.1\pm 0.1\pm0.1\times10^{38}\,{\rm erg}$. Here, the former uncertainty is for the 95\% confidence-level statistical error, while the later one comes from the 20\% $T_{\rm sys}$ variation. 
The natural logarithmic  Bayesian factor ($\ln \mathcal{B}$) of the broken power law model over the single power law model is 1275, which indicates a strong preference for the former. Our measured power-law index at the higher-energy end ($\gamma_2$) is close to that of FRB~20121102A\cite{LiDi2021} (i.e., $\gamma\sim1.4$ for $E>3\times10^{38}\,{\rm erg}$), and it is also close to the power-law index ($\gamma \approx 1.3$) of the bright bursts of FRB~20180916B detected by CHIME\cite{CHIME20Nat}. The power-law index at the lower-energy end ($\gamma_1$) is shallower than that measured in FRB~20121102A\cite{LiDi2021} with $\gamma=0.61\pm0.04$ for $4\times10^{36}<E<3\times10^{38}$. We tested the effects of Gaussian fitting BW on $\gamma$ by comparing the value derived using 2$\sigma$ and 3$\sigma$ BW values, where $\delta\gamma_1=0.02$ and $\delta\gamma_2=0.06$ are comparable to the systematics of $T_{\rm sys}$ variation.} \REV{We searched for the corresponding high-energy transients in both {\it Fermi} and {\it Insight}-HXMT/GECAM data\cite{Zou2021,Cai2021}, and the null detection places loose bounds that the ratio between the luminosity in radio and $\gamma$-ray bands, i.e. $L_{\rm radio} / L_{\rm \gamma}\ge 1.4\times 10^{-7}$ (8--200\,keV) and $\ge 6.3\times 10^{-7}$ (200--3,000\,keV).}

\subsubsection*{Waiting time between the bursts}
The burst times of arrival (TOAs) were measured from the centroid of the best-matched boxcar filter\cite{Men19MN}. We then converted the site arrival times to the barycentric coordinate time at the reference frequency of 1500\,MHz using the software package \textsc{tempo2}\cite{TEMPO2ref}. The waiting times ($\Delta T_{\mathrm{wait}}$) were calculated by subtracting two adjacent barycentric TOAs. The distribution of the waiting time is shown in \EXTFIG{fig:waitingtime}. We modeled it using the superposition of three log-normal distributions, where the best-fitting curve to the histogram is in \EXTFIG{fig:waitingtime}. The waiting-time distribution shows local maxima at 39\,ms, 45.1\,s, and 162.3\,s. We note that a superposition of two log-normal distributions is insufficient to describe the data (\EXTFIG{fig:waitingtime}), which can be verified by {the large value of natural logarithmic Bayesian factor ($\ln \mathcal{B}=35$) of the three-component model over the two-component model.
The multimodal distribution of waiting times is similar to the case of FRB~20121102A\cite{LiDi2021}, except that the shortest waiting-time population (39\,ms) is one order of magnitude longer than that of FRB~20121102A (3.4\,ms)}.

\subsubsection*{Dispersion measure (DM)}

We used the Fourier-domain method to measure the dispersion measure (DM), where the DM is measured by maximising the time derivative of ``intensity'' computed only with the Fourier phase {(\url{https://www.github.com/DanieleMichilli/DM_phase})}. 
The DM values as a function of time are collected in \EXTFIG{fig:radiosup1}.
{The mean value is $413.2\,\cmpc$ and the RMS deviation is $2.0\,\cmpc$. A linear fit shows no obvious trend in DM with a 95\% confidence-level upper limit of} $|{\rm d DM}/{\rm d}t|\le 7\times10^{-3}\,\cmpc\,\mathrm{day^{-1}}$, while there is a peak-to-peak fluctuation of $\sim10\,\cmpc$ in the burst-to-burst DM.

{The measured DM agrees with the theoretical expectation. The observed DM is $\sim 413\,\cmpc$. The Galactic interstellar medium (ISM) contribution is $\DMm \approx 140\,\cmpc$ or $\DMm \approx 200\,\cmpc$ according to the NE2001 model\cite{CL02arXiv} and the YMW16 model\cite{YMW16}, respectively. Considering a Galactic-halo contribution of $\sim30\,\cmpc$\cite{Dolag15MN}, the total Galactic contribution is in the range 170--230\,$\cmpc$. The rest of all the extragalactic DM contributions would thus be 183--243\,$\cmpc$. Further accounting for the intergalactic medium contribution of $\DMi \approx 80\,\cmpc$ at $z=0.09795$\cite{DengZhang2014}, the FRB host-galaxy DM contribution should be in the range 103--163\,$\cmpc$. This value agrees with the modeling of the host galaxy using the DM template technique\cite{Luo18MN}, which predicted that the 68\%-confidence-level range of the host-galaxy DM is $10\le \DMh/(1+z)\le300\,\cmpc$, where the host-galaxy H$\alpha$ luminosity $L_{\rm H\alpha}=7\times10^{41}$\,erg\,s$^{-1}$ and the effective radius $R_{e}=1.5$\,kpc are from our optical observations (see the optical observation section in Methods).}

\subsubsection*{Polarisation properties}\label{sec:polarisation}

Our polarisation data are calibrated with the single-axis model using the software package \textsc{psrchive}\cite{HvSM04}, where both the differential gain and phase between the two polarisation channels are calibrated using noise diode signal injected in the feed. The polarisation fidelity and calibration scheme have been described and tested in previous work\cite{Jiang20RAA,Luo20Nat}. 

We measure the RM for high-quality bursts only with $S/N \ge 30$ (1103 in total) using the $Q-U$ fitting method\cite{DKL19}. We corrected the ionosphere contribution with values computed from the software package \textsc{ionFR}\cite{SSH13}, where the maximal ionosphere RM correction is 3\,$\radm$. The RM in the FRB's source rest frame\cite{WPK84, ARG16} is $\RM_{\mathrm{host}}=(1+z)^2(\RM_{\mathrm{obs}}-\RM_{\mathrm{Gal}})=-380\,\radm$ to $-1010\,\radm$. 

We can exclude a few common origins for the RM variation. (1) The RM variation cannot be explained with the RM contribution in the MW, which is $-51(5)$\,$\radm$ along the direction of FRB~20201124A; the maximal variation is a few tens of radians per square meter {within a $2^\circ$ field of view}\cite{XuHan14}. (2) The RM variation is not caused by instrumental artifacts. In polarisation studies, we have excluded the data of April 1 and 2, where the observations were carried out with off-axis illumination. The FAST polarimetry stability has been checked\cite{Luo20Nat} to show that the RM measurement is stable with $\Delta{\rm RM} \le 0.2\,\radm$. We also checked if saturation or nonlinearity affected our polarimetry. The digital sampling and data recording is done with an 8-bit sampling scheme at FAST. We tested the nonlinearity by comparing the differences in results between including and removing the data above 250 (the maximal digital value is 255 for an 8-bit system); the differences are tiny. (3) RM variation is not from the apparent RM variation across the phase of a burst as found in pulsars\cite{NKK09} and in FRBs\cite{Cho20}. We find that the maximum amplitude of RM variations within single bursts for FRB 20201124A is at the level of 15\,$\radm$ (examples are shown in \EXTFIG{fig:burstrm}), which is too small to explain the RM variation of FRB 20201124A at the level of $\sim 500$\,$\radm$. (4) RM variation is not caused by the intrinsic frequency evolution of individual bursts, as the PA rotation matches the cold plasma Faraday rotation relation. We relaxed the power-law index of wavelength in $Q-U$ fitting and also fit for the RM index $\beta$ using the model $\Delta\Psi=\mathrm{RM}\,\lambda^\beta$. One expects $\beta=2$ if the cold plasma Faraday rotation model can be applied, while the index $\beta$ would not necessarily equal 2 if the apparent RM is caused by intrinsic profile evolution. As shown in \EXTFIG{fig:RMindex}, we found that for 57\% of the bursts (631 out of 1103 bursts), the deviations of measured RM index values are within 1$\sigma$ error bars. Visual inspection revealed that the $\beta \neq 2$ deviation was mainly caused by overlapping of multiple components in the dynamic spectrum. The detected trend of RM variation is hardly affected by the $\beta \neq 2$ deviation as shown in \EXTFIG{fig:RMindex}. To further reduce the profile-evolution effects, only the measurements with RM index within 1$\sigma$ of $\beta = 2$ are included in \FIG{fig:fig1}.

Considering the 10-day timescale of RM variation, we expect that the major RM variation comes from the FRB local environment, {over a distance scale of $\Delta X \approx 0.6\, {\rm au}\, (\tau/10\,{\rm day}) (v/100\,{\rm km\,s^{-1}})$, where we normalised velocity to the typical value of binary motion or proper motion of neutron stars.} {We can derive a bound on the parallel magnetic field by using the constraint induced by free-free absorption at such a small distance scale.
The detection of FRB emission requires that the free-free absorption optical depth should be smaller than unity\cite{MH67,Hessels19ApJL}, which means that the local DM contributing to the RM variation meets
\begin{equation}
    {\rm \Delta DM}< 4 \eta \left(\frac{T}{10^4}\right)^{0.7}\left(\frac{\nu}{\rm GHz}\right)^{1.1} \left(\frac{\Delta X}{\rm au}\right)\,,
\end{equation}
where we converted the DM to EM via the filling factor $\eta\equiv\frac{\Delta {\rm DM}^2}{\Delta X {\rm EM} }\le 1$. One finds
\begin{equation}\langle B_{\|}\rangle= 1.23~\mathrm{\mu G}\,\frac{\Delta \RM}{\Delta \DM} > 0.2~\mathrm{mG} \, \eta^{-1} \left(\frac{\Delta \RM}{\rm 500\, rad\, m^{-2}}\right)\left(\frac{T}{10^4}\right)^{-0.7}\left(\frac{\nu}{\rm GHz}\right)^{-1.1} \left(\frac{\Delta X}{\rm au}\right)^{-1}\,.
\end{equation} 
\REV{Here RM is defined in the source rest frame.}
Such a mG-level lower bound of the magnetic field was also reported for FRB~20121102A\cite{Michilli18Nat,VedanthamRavi2019} based on the host-galaxy DM estimation. }

For 41 of the total 1863 bursts, we have discovered {oscillations of fractional circular/linear polarisation ($\Pi_{\rm V}\equiv V/I, \Pi_{\rm L}\equiv L/I$), and linear PA}. {The oscillation can be characterised with periodicity in the square of wavelength, and we use the Lomb-Scargle periodogram\cite{Lomb1976} to find such features}. A $\chi^2$ fitting to the following model simultaneously for circular and linear polarisation intensities is used to infer the oscillation parameters:
{
\begin{eqnarray}
L&=&I \left[\Pi_{\rm L0}+\dot{\Pi}_{\rm L} \lambda^2+A\sin(\omega_{\lambda^2}\lambda^2+\phi_{\rm L})\right]\,,\label{eq:fitmod1}\\
V&=&I \left[\Pi_{\rm V0}+\dot{\Pi}_{\rm V} \lambda^2+A\sin(\omega_{\lambda^2}\lambda^2+\phi_{\rm V})]\right]\,,\label{eq:fitmod2}\\
Q&=&L \cos \left[ {\rm PA}_0+A_{\rm PA}\cos(\omega_{\lambda^2}\lambda^2+\phi_{\rm PA}) \right]\,,\label{eq:fitmodq}\\
U&=&L \sin \left[ {\rm PA}_0+A_{\rm PA}\cos(\omega_{\lambda^2}\lambda^2+\phi_{\rm PA}) \right]\,,\label{eq:fitmodu}
\end{eqnarray}
where the parameters $\Pi_{\rm L0}$, $\dot{\Pi}_{\rm L}$, $\Pi_{\rm V0}$, and $\dot{\Pi}_{\rm V}$ are the average values and the slopes of the fractional linear and circular polarisation, $A$ is the amplitude of oscillation in the fractional polarisation, $\rm PA_{0}$ is the mean PA, and $A_{\rm PA}$ is the amplitude of the oscillation in PA. A common angular frequency of oscillation ($\omega_{\lambda^2}$) is assumed for $\Pi_{\rm V}$,$\Pi_{\rm L}$ and PA. The $\phi_{\rm L}$,  $\phi_{\rm V}$, and $\phi_{\rm PA}$ are oscillation phases.}  The best-fitting conjugate frequencies of bursts 779 and 926 are $\omega_{\lambda^2}=2400\pm30 \,\radm$ and $1800\pm10\,\radm$. In the framework of mild polarisation absorption or Faraday conversion, one has $|{\rm RM'}|=\omega_{\lambda^2}/2$, {which is the Faraday rotation accumulated from the FRB, along the line of sight, up to the position where the absorption or conversion occurs}: ${\rm RM}' = 1200\pm15 \,\radm$, $900\pm5\,\radm$, respectively. {We note that the total observed RM is of the same order of magnitude, which indicates that a significant amount of RM comes from the vicinity of the FRB source, where, at the same time, Faraday conversion or synchrotron absorption is still important\cite{Sazonov69,HuangShcherbakov2011}, i.e. a cold and a relativistic plasma coexist.}

We plot the best-fit curves against the data in \FIG{fig:fig2}, where we convert the model to fractional polarisation for better visualisation. The best-fit phase differences between the linear and circular oscillations are given in panel (d) of \FIG{fig:fig2}. For burst 779, the oscillations of $\Pi_{\rm V}$ and $\Pi_{\rm L}$ decrease significantly above 1160\,MHz ($\lambda^2\leqslant 0.067\,{\rm m}^2$ as indicated by the shaded area in \FIG{fig:fig2}), where the best-fit amplitudes of oscillation below and above 1160\,MHz are $0.16\pm0.01$ and $0.008\pm0.005$, respectively. Such frequency-dependent oscillation can be explained with the characteristic frequency of a uniformly magnetised plasma\cite{HuangShcherbakov2011} that the plasma effects are important when wave frequency bellow the characteristic frequency, i.e. $f\le0.04{\,\rm GHz}\,(B/{\rm G}) \gamma^2$, with $\gamma$ the electron kinetic Lorentz factor. In the polarisation-dependent radiative transfer framework, the constraint for magnetic field becomes $B\ge 3\,{\rm G} (f/{\rm 1.1 GHz})(\gamma/10)^{-2}$.

We checked the power-law index of oscillation with respect to $\lambda$ by replacing terms of $\omega_{\lambda^2}\lambda^2$ in \EQ{eq:fitmod1} and (\ref{eq:fitmod2}) to a generalised form of $\omega_{\lambda^{2k}} \lambda^{2k}$ and fit the index $k$ simultaneously. For bursts 779 and 926, we had $k=0.998\pm0.005$ and $1.0\pm0.1$, which verifies the $\lambda^2$-dependent oscillation of polarisation.

Besides Faraday conversion, polarisation-dependent scintillation can also induce such $\lambda^2$-dependent oscillations\cite{Beniamini2021}, {if the number of slits on the scintillation screen is limited. However, special conditions are required in this model to reproduce the observation. First, polarisation-dependent absorption is still required to induce oscillation in total polarisation $\Pi_{\rm P}$. Second, contrived fine tuning is required to achieve a 40\% $\Pi_{\rm V}$ variation for a $<10^\circ$ PA change as seen in burst 779. Also, an extra mechanism is needed to cease the polarisation oscillations above 1160\,MHz for burst 779. }

We note that not all bursts with the measured nonzero $\Pi_{\rm V}$ show the above oscillatory behaviour. Some bursts exhibit slow variations with opposite phases of $\Pi_{\rm V}$ and $\Pi_{\rm L}$, such as burst 1112 in \FIG{fig:fig2}. The variation may come from Faraday conversion or an intrinsic radiation mechanism of FRBs. Interestingly, the burst with the highest $\Pi_{\rm V}$ in our sample (burst 1472 in \FIG{fig:fig2}) shows no significant oscillation. Therefore, on top of Faraday conversion, an alternative, intrinsic radiation mechanism may be required to generate circular polarisation.

\subsubsection*{Keck optical and near-infrared observations}

We obtained Keck\,I low- and high-dispersion spectra with the 
Low Resolution Imaging Spectrometer (LRIS)\cite{Oke1995, Rockosi2010}
and the Echellette Spectrograph Imager (ESI)\cite{Sheinis2002}, respectively.
The LRIS spectra were taken with a slit width of $1.0\arcsec$ at PA = $53.4^\circ$, and there were $750 + 920$\,s and $2 \times 750$\,s exposures on 
the blue and red sides, respectively. LRIS has an atmospheric dispersion 
corrector. The data were reduced using LPipe \cite{Perley2019}, and the fluxes 
were scaled to match Pan-STARRS1 \cite{Pan-STARRS12020} $griz$ photometry.  
Galactic extinction corrections\cite{Cardelli1989, Schlafly2011} were applied 
with $R_V=3.1$ and $E(B-V)_{\rm MW} = 0.652$\,mag.
We took $8 \times 320$\,s exposures with ESI in the cross-dispersed echelle mode with resolving power $R\approx 10,000$ and a slit width of $1.0\arcsec$ at the parallactic angle\cite{Filippenko1982} of $87^\circ$. They were reduced with  ESIRedux  {(\url{https://www2.keck.hawaii.edu/inst/esi/ESIRedux/index.html})}, with only relative-flux calibration performed.

The LRIS imaging consisted of $4 \times 180$\,s exposures in the $g$ band and $2 \times 180$\,s in the $i$ band. They were reduced following standard procedures of bias subtraction, flat fielding, and coadding.

We obtained $4 \times 120$\,s near-infrared $K^\prime$-band images
(dithered by 3--4$\arcsec$ between exposures) with the NIRC2 camera
($0.04\arcsec$\,pixel$^{-1}$ scale and 
$40\arcsec$\ field) via the Keck\,II laser-guide-star AO system \cite{Wizinowich06}. An $R = 15.9$\,mag star $36\arcsec$\ NW of the FRB host 
served as the tip-tilt reference star. The images were reduced 
following a standard iterative procedure implemented in an IDL package {(\url{https://github.com/fuhaiastro/nirc2})}, and
the final combined image reaches a full width at half-maximum intensity (FWHM) resolution of $0.12\arcsec$. 
{The astrometry is calibrated using five stars in the field to the {\it Gaia} reference frame (Gaia-CRF2 \cite{Gaia-CRF2}), which is tied to the International Celestial Reference System (ICRF) reference system using accurate VLBI positions of quasars at sub-mas precision.}

\subsubsection*{Properties of the host galaxy} 

\paragraph*{Star-formation rate and gas-phase metallicity:} We use the emission lines detected in the high-$S/N$ LRIS spectrum to {measure the redshift ($z=0.09795\pm0.00003$)}, infer the star-formation rate (SFR), and determine the gas-phase metallicity of the galaxy. We measure line fluxes of ${\rm H}\alpha$, $\ion{\rm N}{ii}\,\lambda6548$, $\ion{\rm N}{ii}\,\lambda6583$, $\ion{{\rm O}}{iii}\,\lambda5007$, $\ion{{\rm O}}{ii}\,\lambda3726$,\,3729, and $\ion{\rm S}{ii}\, \lambda6716$,\,6731 by fitting single-Gaussian profiles; for H$\beta$, we add an additional Lorentzian component to account for stellar absorption. 

We use the ${\rm H}\alpha$ luminosity $L({\rm H}\alpha)$ to estimate the SFR. The internal extinction of the galaxy is estimated with the Balmer decrement by adopting $({\rm H}\alpha/{\rm H}\beta)_{\rm theory}=2.86$ for Case B recombination and using the Calzetti et al.\cite{Calzetti2000} reddening curve; we obtain ${E}(B-V)=0.43\pm0.04$\,mag and thus ${A}_\lambda({\rm H}\alpha)= 1.27\pm0.12$\,mag, yielding $L({\rm H}\alpha) = (6.9\pm0.7)\times10^{41}$\,erg\,s$^{-1}$, which translates to SFR $= 3.4\pm0.3\,M_{\odot}$\,yr$^{-1}$\cite{Heintz2020}. Our SFR estimate is higher than previous $L({\rm H}\alpha)$-based measurements, $\sim 2.1\,{\rm M_{\odot} yr^{-1}}$ \cite{Fong21}, $\sim 1.7\,{\rm M_{\odot} yr^{-1}}$  \cite{Ravi21}, and $2.3\pm0.4\,{\rm M_{\odot} yr^{-1}}$ \cite{Piro21}, while it is lower than those derived from spectral energy distribution (SED) fitting ($\sim 4.3\,{\rm M_{\odot} yr^{-1}}$ \cite{Fong21}) and radio data ($\sim 7\,{\rm M_{\odot} yr^{-1}}$ \cite{Ravi21} and $\sim 10\,{\rm M_{\odot} yr^{-1}}$ \cite{Piro21}). Adopting a stellar mass $M_* = (2.5\pm0.7) \times10^{10}\,M_{\odot}$ for the galaxy from averaging two existing results\cite{Fong21, Ravi21}, the specific SFR is $\log({\rm sSFR/{\rm yr}^{-1}}) = -9.86\pm0.11$. 
We also cross-check it by estimating sSFR using the equivalent width (EW) of H$\alpha$, EW(H$\alpha$) = 48\,\AA\, and obtain $\log({\rm sSFR/{\rm yr}^{-1}}) = -9.65\pm0.19$ \cite{Belfiore2018}; it is higher than our $L({\rm H}\alpha)$-based estimate by $\sim 1\,\sigma$. 

We infer the gas-phase metallicity ($Z$) by applying the Inferring metallicities ($Z$) and Ionization parameters ($q$) (IZI) photoionisation model\cite{Blanc2015,Mingozzi2020} to the fluxes of all the above-mentioned emission lines,
yielding a best-fit oxygen abundance of 12 + log(O/H) = $9.07^{+0.03}_{-0.04}$, in agreement with a previous estimate\cite{Fong21} (12 + log(O/H) = $9.03^{+0.15}_{-0.24}$) using the ``O3N2'' method.

\paragraph*{Morphology and kinematics:}
The left and middle subpanels of \EXTFIG{fig:opticalobs01}(a) show the LRIS $i$-band image and the NIRC2 $K^\prime$-band AO image, respectively. 
The AO image with FWHM $= 0.12\arcsec$ enables us to resolve the bar and spiral features of the galaxy, which is not possible with natural seeing. We used GALFIT \cite{Peng2010} to model the host galaxy in the NIRC2 image with a single-component model composed of a S\'ersic profile in the radial direction and a generalised ellipse function in the azimuthal direction. We obtain the best-fit effective radius $R_e = 1.5$\,kpc and axis ratio $b/a=0.62$, which suggests cos\,$(i)$ = 0.6 (where $i$ is disc inclination angle) \cite{Holmberg1958, TF1977AA}.

After subtracting the disc component, the galaxy bar and spiral features can be clearly seen in the residual NIRC2 image shown in the right-most subpanel of \EXTFIG{fig:opticalobs01}(a). We  measure the centroid of the galaxy bar by fitting a two-dimensional (2D) Gaussian model and obtain refined J2000 coordinates of the galaxy centre {(RA $= 05^{\rm h}08^{\rm m}03.4896^{\rm s}$, Dec $= +26^\circ03^\prime37.869^{\prime\prime}$). The FRB is $0.239\pm0.013\arcsec$ and $0.636\pm0.007\arcsec$ to the east and north of the galaxy centre, respectively. The uncertainties are estimated by adding the localisation uncertainties from EVN ($\Delta{\rm RA}_{\rm EVN}=4.5$\,mas and $\Delta{\rm Dec}_{\rm EVN}=3.6$\,mas \cite{FRB20201124AEVN}) and optical observations ($\Delta{\rm RA}_{\rm opt}=12.0$\,mas and $\Delta{\rm Dec}_{\rm opt}=6.5$\,mas) in quadrature. The error budget of optical astrometry is dominated by the uncertainties from the Keck-{\it Gaia} frame transformation, whereas the centroiding errors in Keck (2\,mas) and {\it Gaia} (from a fraction of 1\,mas to $\sim2$\,mas for the five reference stars) are relatively minor.

The object's projected position is on the disc, although it does not seem to coincide with any other visible structures. We measure the one-dimensional intensity profile of a chord slicing through the FRB position and perpendicular to the major axis of the bar on the NIR2 image. We find that it is well described by a model of concentric double Gaussian functions with the broader ($\sigma=511$\,mas) and narrower ($\sigma=96$\,mas) Gaussians describing the disc and the bar contributions, respectively. We subtract the broader Gaussian (i.e., the disc contribution) and find that at the position of the FRB, which is $260\pm13$\,mas from the bar centre, its light has only $7\%$ of the peak bar intensity, which is very small.}

As shown in \EXTFIG{fig:opticalobs01}(b), the H$\alpha$ line in the ESI spectrum has a double-peaked profile with a peak-to-peak separation of $\sim 100$\,km\,s$^{-1}$, which may be due to disc rotation; however, since the ESI slit was oriented along the minor axis of the galaxy, it may alternatively be caused by gas outflow. We study the disc rotation with LRIS, for which the slit was oriented $60^\circ$ with respect to the major axis. As shown in the left subpanels of \EXTFIG{fig:opticalobs01}(c), the wavelength centroids of H$\alpha$ emission vary along the LRIS slit direction. We extract H$\alpha$ lines with a step size of 3 pixels ($0.4\arcsec$) along the slit direction, and we measure their projected galactocentric distance ($r_\perp$) and line-of-sight velocities ($v$) to the continuum centre shown as the black dots in the right subpanel of \EXTFIG{fig:opticalobs01}(c). Then we fit the data using a simple rotational disc model, in which velocity scales linearly with galactocentric distance $r$ for $r<r_{\rm break}$ (the velocity zero point is a free parameter) and stays constant at $v_{\rm ROT}$ for $r>r_{\rm break}$. The best-fit model, which is shown as the red line in the right subpanel of \EXTFIG{fig:opticalobs01}(c), has the deprojected rotation velocity $v_{\rm ROT}=139\pm 19\,{\rm km\,s^{-1}}$ and $r_{\rm break}=3.0\pm0.5$\,kpc. Our $v_{\rm ROT}$ estimate suggests a galaxy stellar mass $M_* \approx 2\times10^{10}\,M_\odot$ using the Tully-Fisher relation \cite{TF1977AA, Ouellette2017}. This is consistent with our adopted value $M_* = (2.5\pm0.7) \times 10^{10}\,M_{\odot}$ from averaging two previous estimates \cite{Fong21, Ravi21}.

\subsubsection*{{Background galaxy}}

Fong et al.\cite{Fong21} tentatively identified a background galaxy with the possible detection of H$\beta$ and \ion{{\rm O}}{iii} emission lines at $z = 0.5531$. Our spectra allow its firm identification and study of its properties. We detect H$\alpha$, H$\beta$, and \ion{{\rm O}}{iii}\,$\lambda\lambda4959$,\,5007 emission lines at $z=0.5534\pm0.0001$ in the LRIS (blue) and ESI (cyan) spectra (see \EXTFIG{fig:opticalobs02}(c)). Owing to the nondetection of \ion{{\rm N}}{ii} (or \ion{{\rm O}}{ii}), we cannot distinguish between a star-forming galaxy and an active galactic nucleus (AGN) in the Baldwin-Phillips-Terlevich (BPT) diagram (see \EXTFIG{fig:opticalobs02}(a)). The \ion{{\rm O}}{iii}\,$\lambda$5007 line is resolved by ESI with a velocity dispersion $\sigma_{{\ion{{\rm O}}{iii}}\,\lambda5007} = 27.6 \pm 2.6$\,km\,s$^{-1}$;
such a low velocity dispersion  favours that it is a star-forming galaxy\cite{Law2021}. Using IZI, we find that its gas-phase metallicity 12 + log(O/H) $= 8.29^{+0.26}_{-0.28}$ and  $E(B-V)=0.27^{+0.12}_{-0.13}$\,mag.  The extinction-corrected ${\rm H}\alpha$ luminosity is $L({\rm H}\alpha) = 1.14^{+0.51}_{-0.38}\times10^{42}$\,erg\,s$^{-1}$, which yields SFR = $5.7^{+2.5}_{-1.9}\,M_{\odot}$\,yr$^{-1}$. 

The centres of \ion{{\rm O}}{iii} $\lambda5007$ emission from the background galaxy are offset from the centre of the continuum dominated by the host (foreground) galaxy in the 2D spectroscopic image.  We measure the centroid of the background galaxy by using its \ion{O}{iii} emission line detected at two slit orientations, which are shown in \FIG{fig:fig3}(b); it is $0.29\arcsec$ west and $0.22\arcsec$ north of foreground galaxy's centre. Owing to the large projected separation of 4.7\,kpc, and inconsistency of the expected DM and scattering timescale\cite{Main2021}, it is unlikely that the background galaxy is the FRB host.

\clearpage

\subsubsection*{Data availability}
Raw data are available from the FAST data center, \url{http://fast.bao.ac.cn}. Owing to the large data volume, we encourage contacting the corresponding author for the data transfer. The directly related data that support the findings of this study can be found at the PSRPKU website, \url{https://psr.pku.edu.cn/index.php/publications/frb20201124a/} and Figshare website, \url{https://doi.org/10.6084/m9.figshare.19688854}. 

\subsubsection*{Code availability}
\noindent
\textsc{PSRCHIVE} (\url{http://psrchive.sourceforge.net})

\noindent
\textsc{transientx} (\url{https://github.com/ypmen/TransientX})

\noindent
\textsc{BEAR} (\url{https://psr.pku.edu.cn/index.php/publications/frb180301/})

\begin{addendum}

\item 
We are grateful to L. C. Ho, H. Gao, and R. Li for discussions.
This work made use of data from the FAST. FAST is a Chinese
national megascience facility, built and operated by the National
Astronomical Observatories, Chinese Academy of Sciences.  We acknowledge the use of public data from the Fermi Science Support Center (FSSC). This work is supported by National SKA Program of China (2020SKA0120100, 2020SKA0120200), Natural Science Foundation of China (12041304, 11873067, 11988101, 12041303, 11725313, 11725314, 11833003, 12003028, 12041306, 12103089, U2031209, U2038105, U1831207), National Program on Key Research and Development Project (2019YFA0405100, 2017YFA0402602, 2018YFA0404204, 2016YFA0400801), Key Research Program of the CAS (QYZDJ-SSW-SLH021), Natural Science Foundation of Jiangsu Province (BK20211000), Cultivation Project for FAST Scientific Payoff and Research Achievement of CAMS-CAS, the Strategic Priority Research Program on Space Science, the Chinese Academy of Sciences (grants XDA15360000, XDA15052700, XDB23040400), funding from the  Max-Planck Partner Group, the science research grants from the China Manned Space Project (CMS-CSST-2021-B11,NO. CMS-CSST-2021-A11), and PKU development grant 7101502590. AVF's group at U.C.~Berkeley is supported by the Christopher R. Redlich Fund, the Miller Institute for Basic Research in Science (in which AVF was a Miller Senior Fellow), and many individual donors. SD acknowledges support from the XPLORER PRIZE. BBZ is supported by Fundamental Research Funds for the Central Universities (14380046), and the Program for Innovative Talents, Entrepreneur in Jiangsu. Some of the data presented herein were obtained at the W.M. Keck Observatory, which is operated as a scientific partnership among the California Institute of Technology, the University of California, and NASA; the observatory was made possible by the financial support of the W.M. Keck Foundation. We thank the Keck staff for their help during the observing runs.

\item[Author Contributions] 
HX, JRN, and PC led the data analysis. KJL, WWZ, SD, and BZ coordinated the observational campaign, cosupervised data analyses and interpretations, and led the paper writing. JCJ conducted the polarisation and RM measurements. BJW, JWX, CFZ, and KJL did the timing analysis, periodicity search, DM measurement, burst searching, and Faraday conversion measurement. YPM contributed to the searching software development. RNC, MZC, LFH, YXH, ZYL, ZXL, YHX, and JPY performed software testing. DJZ, YKZ, PW, YF, CHN, FYW, XFW, and SBZ contributed to radio data analysis. PC, SD, HF, AVF, EWP, TGB, SGD, PG, DS, AS, WKZ, and AE contributed to the optical observations and data reduction; AVF also edited the manuscript in detail. PC, SD, HF, and YL contributed to analysing and interpreting the optical data. PJ, HQG, JLH, JLH, HL, LQ, JHS, RY, YLY, DJY, and YZ aided with FAST observations. JLH, DL, MW, and NW helped with observation coordination. KJL, BZ, DZL, WYW, RXX, WL, YPY,WFY, ZGD, and RL provided theoretical discussions. CC, CKL, XQL, WXP, LMS, SX, SLX, JY, XY, QBY, BBZ, SNZ, and JHZ contributed to the high-energy observations and data analyses.

 \item[Competing Interests] The authors declare no
competing financial interests.

\item[Correspondence] Requests for materials should be addressed to the following: \\
K.~J.~Lee (E-mail:kjlee@pku.edu.cn)\\
S.~Dong (E-mail: dongsubo@pku.edu.cn)\\
W.~W.~Zhu (E-mail:zhuww@nao.cas.cn)\\
B.~Zhang (Email:bing.zhang@unlv.edu)\\

\end{addendum}

\clearpage

\clearpage

\setcounter{figure}{0}
\setcounter{table}{0}
\captionsetup[table]{name={\bf Extended Data Table}}
\captionsetup[figure]{name={\bf Extended Data Figure}}

\begin{figure}
\centering
\includegraphics[width=\textwidth]{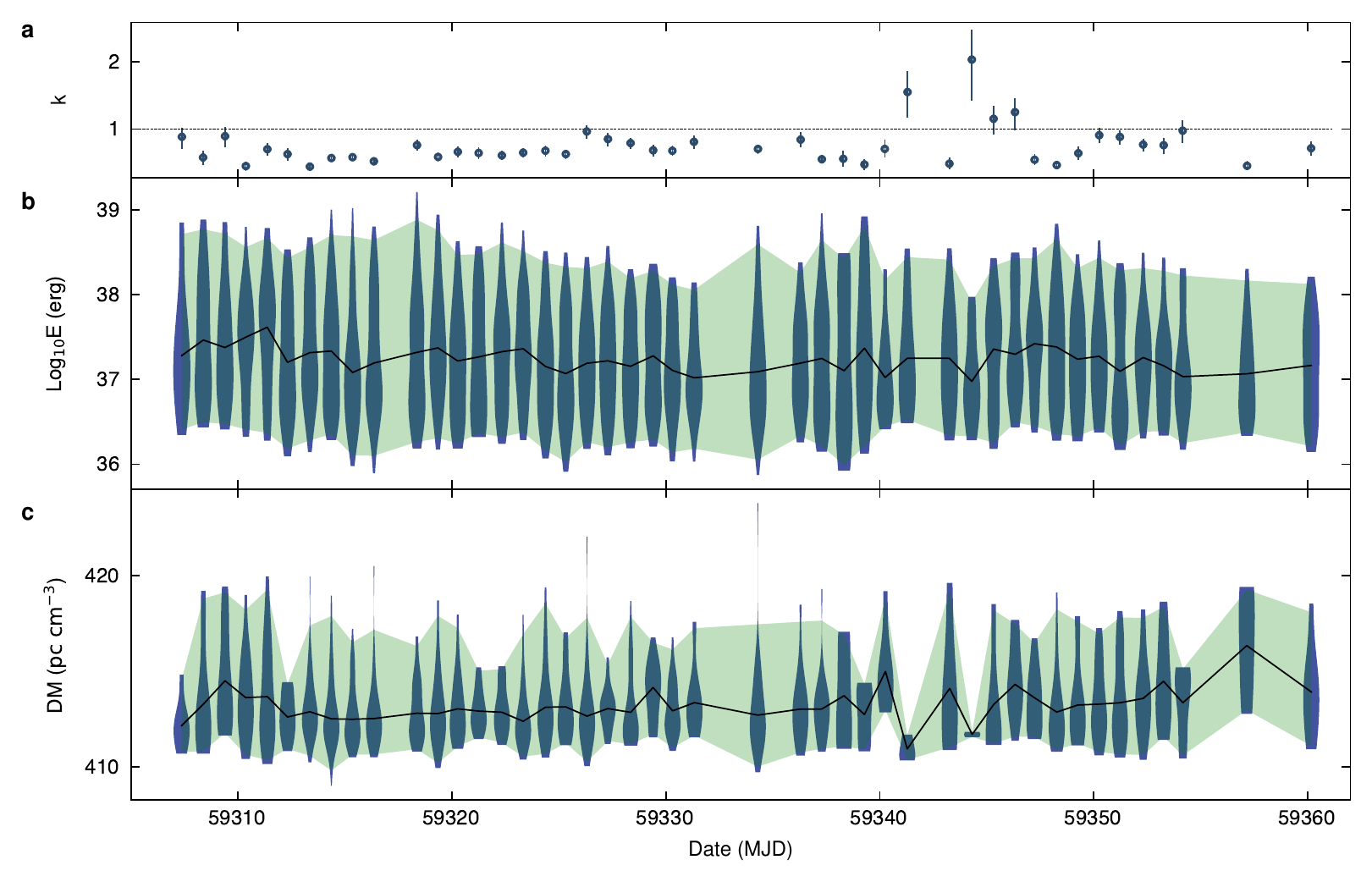}
\caption{{\textbf{Temporal variations of extra physical parameters.} {\bf (a)} Shape parameter ($k$) of Weibull distribution in event-rate inference. The error bar is at 68\% confidence level. {\bf (b)} and {\bf (c)} Daily burst energy and DM, where the violin symbol indicates the distribution function, the green shaded strips indicate the 95\% upper and lower bounds, and the solid black curve is the median. 
}
\label{fig:radiosup1} }
\end{figure}

\begin{figure} 
\centering
\includegraphics[width=\textwidth]{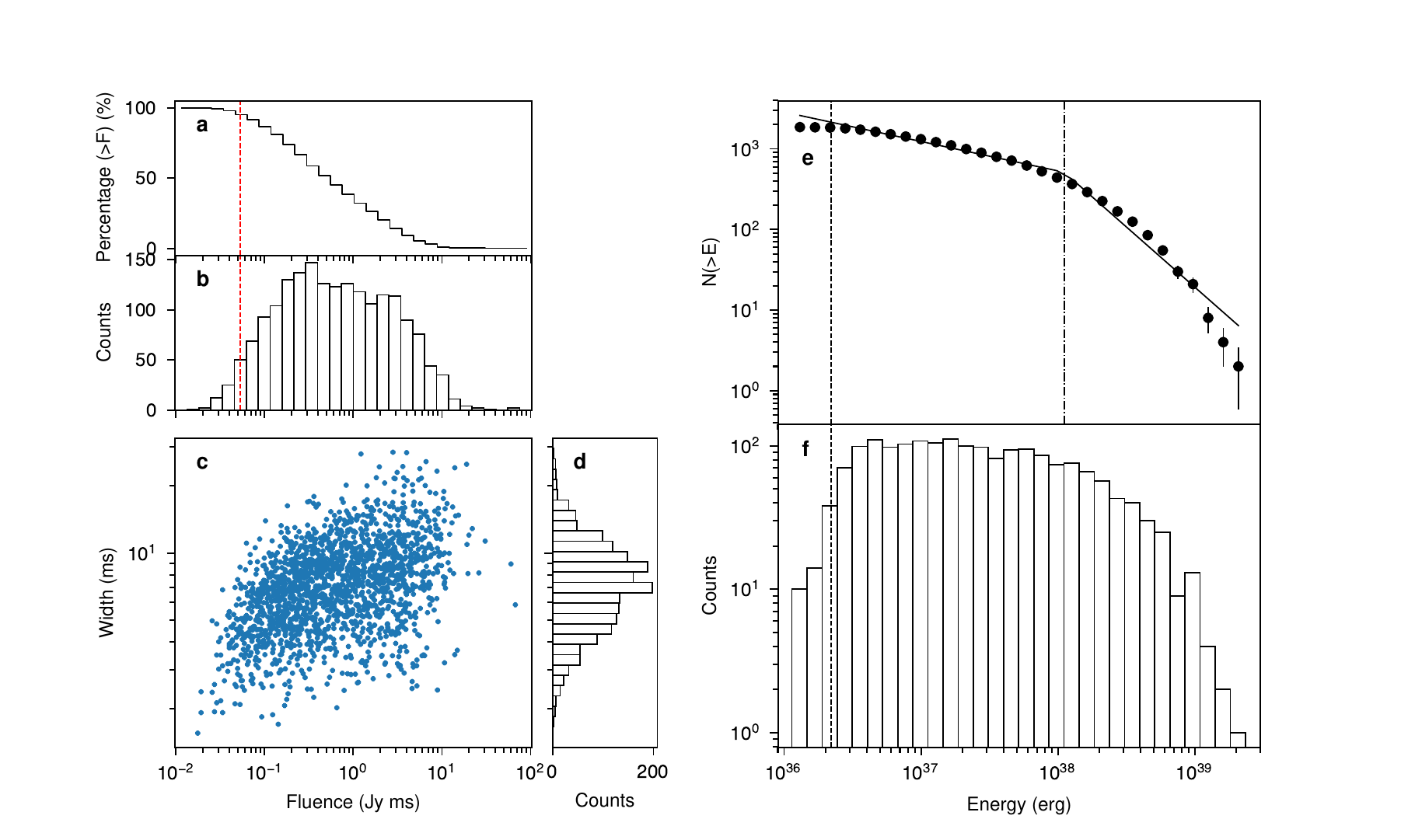}
\caption{{\textbf{Fluence, equivalent width, and energy distribution for detected bursts.}} {\bf (a)} and {\bf (b)} Cumulative distribution and histogram of the burst fluence; the red dashed vertical line at 53\,mJy\,ms indicates the 95\% completeness threshold. {\bf (c)} The 2D distribution of fluence and burst width. {\bf (d)} Histogram of burst width. {\bf (e)} and {\bf (f)} Cumulative distribution and histogram of FRB~20201124A burst energy; {the black dashed vertical line at $2\times 10^{36}\,{\rm erg}$ indicates 95\% completeness assuming a burst bandwidth of 185\,MHz, the median of the burst bandwidths. The broken power-law fit to the cumulative distribution of energy is the solid black curve, with the break point at $1.1\times10^{38}\,{\rm erg}$ indicated by a dot-dashed vertical line.}
\label{fig:flux} }
\end{figure}

\begin{figure} 
\centering
\includegraphics[width=1.0\textwidth]{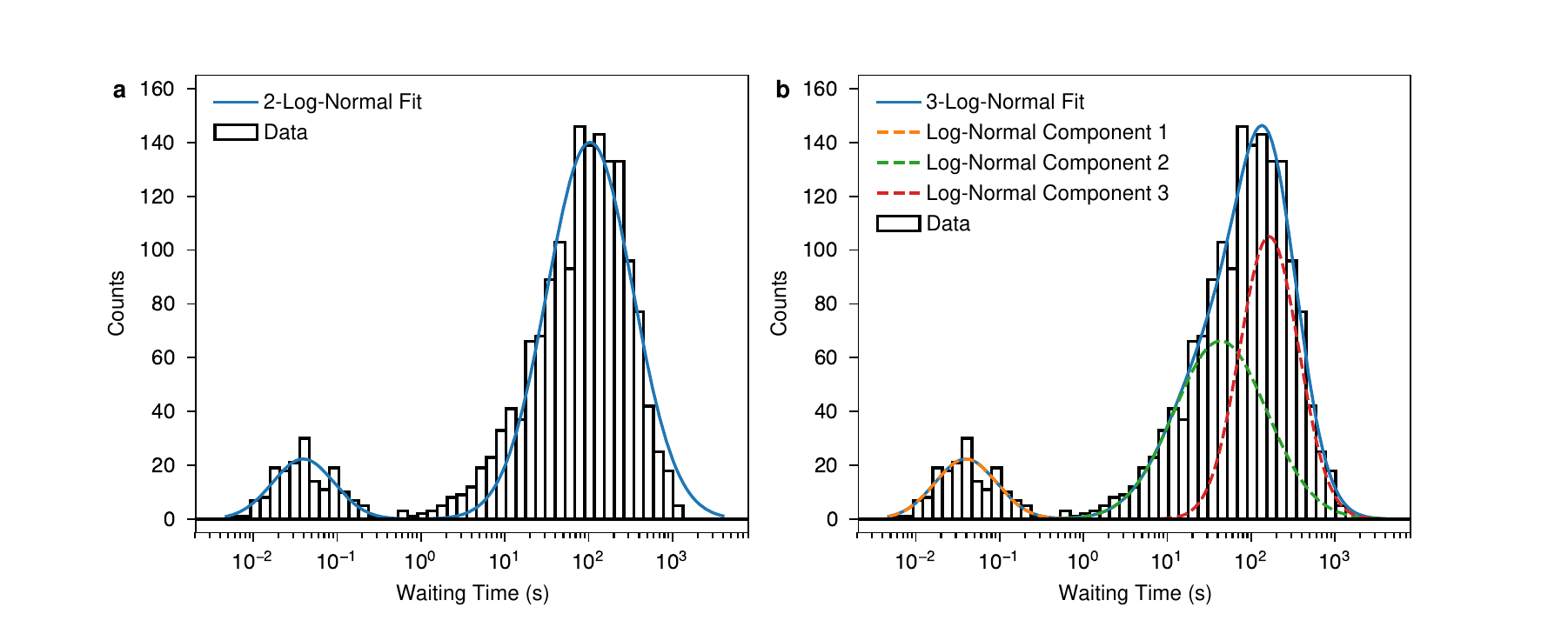}
\caption{{\textbf{Waiting time distribution of FRB~20201124A.}}  {\bf (a)} The best fit using two log-normal functions (the blue curve), where the two log-normal distributions peak at 39\,ms and 106.7\,s. {\bf (b)} The best fit (blue curve) using three log-normal functions, which were indicated with the dashed-line curves, peak at 39\,ms, 45.1\,s, and 162.3\,s.
\label{fig:waitingtime} }
\end{figure}

\begin{figure} 
\centering
\includegraphics[width=1.0\textwidth]{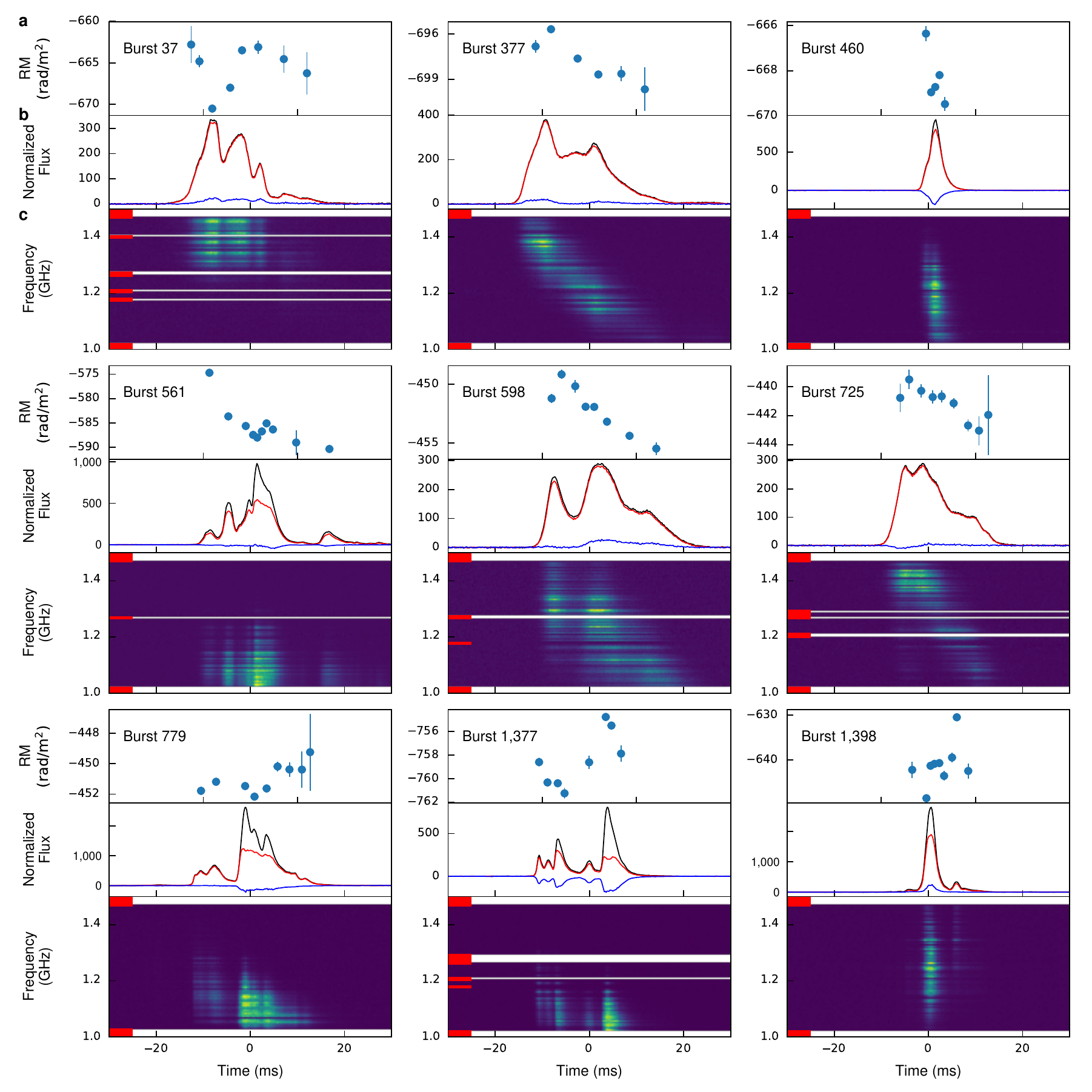}
\caption{{\textbf{Apparent RM variation within individual bursts.}} {\bf (a)} RM curve with 95\% confidence level error bars, {\bf (b)} polarisation profiles, and {\bf (c)} dynamic spectra. Bursts are dedispersed using corresponding structure-optimised DM values. \label{fig:burstrm}}
\end{figure}

\begin{figure} 
\centering
\includegraphics[width=\textwidth]{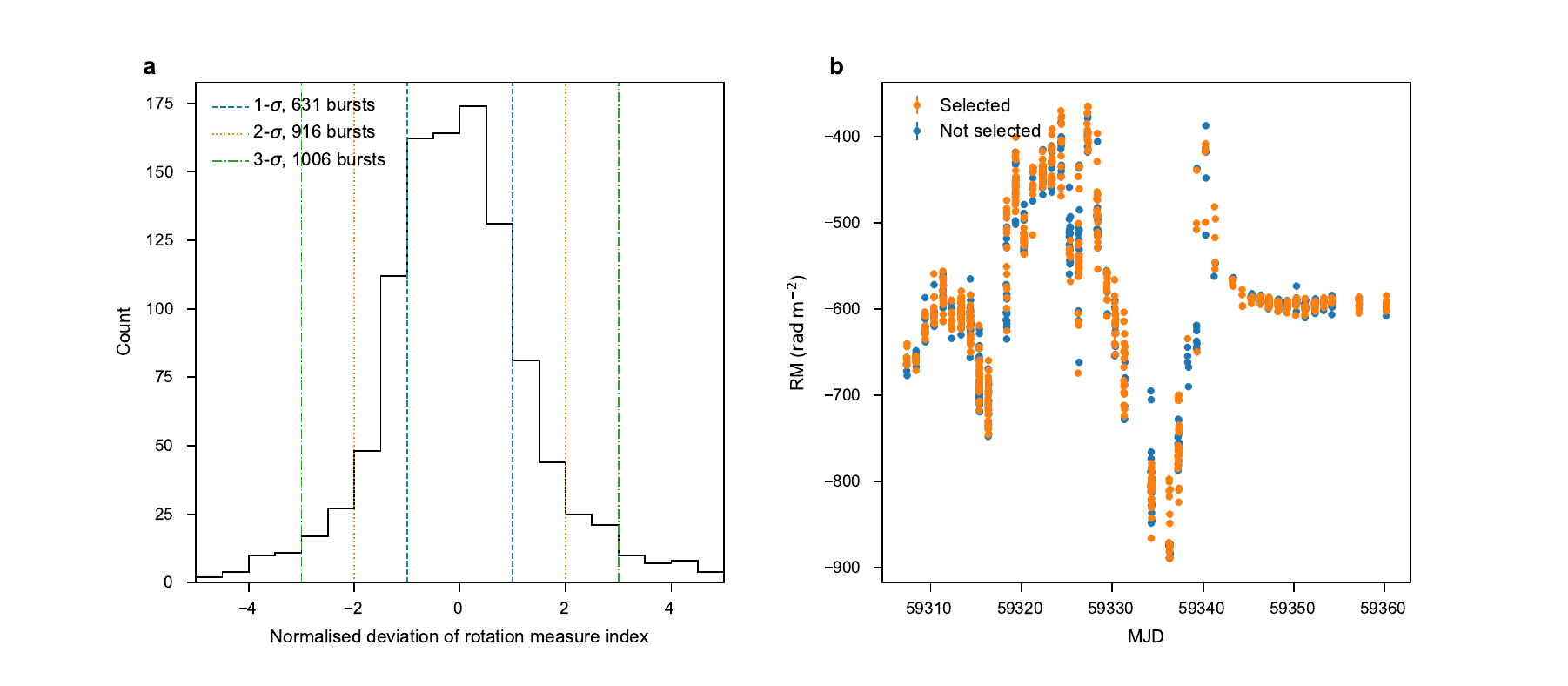}
\caption{\textbf{Rotation measure index.} {\bf (a)} Histogram of normalised rotation measure index deviation defined as $(\beta-2)/\sigma_\beta$, where $\sigma_\beta$ is the uncertainty of $\beta$ with 68\% confidence level. {\bf (b)} RM as a function of time. Orange dots are for selected bursts with $(\beta-2)/\sigma_\beta \leqslant 1$, and the measurements not selected are in blue dots. 
\label{fig:RMindex}}
\end{figure}

\begin{figure} 
\centering
\includegraphics[width=1.0\textwidth]{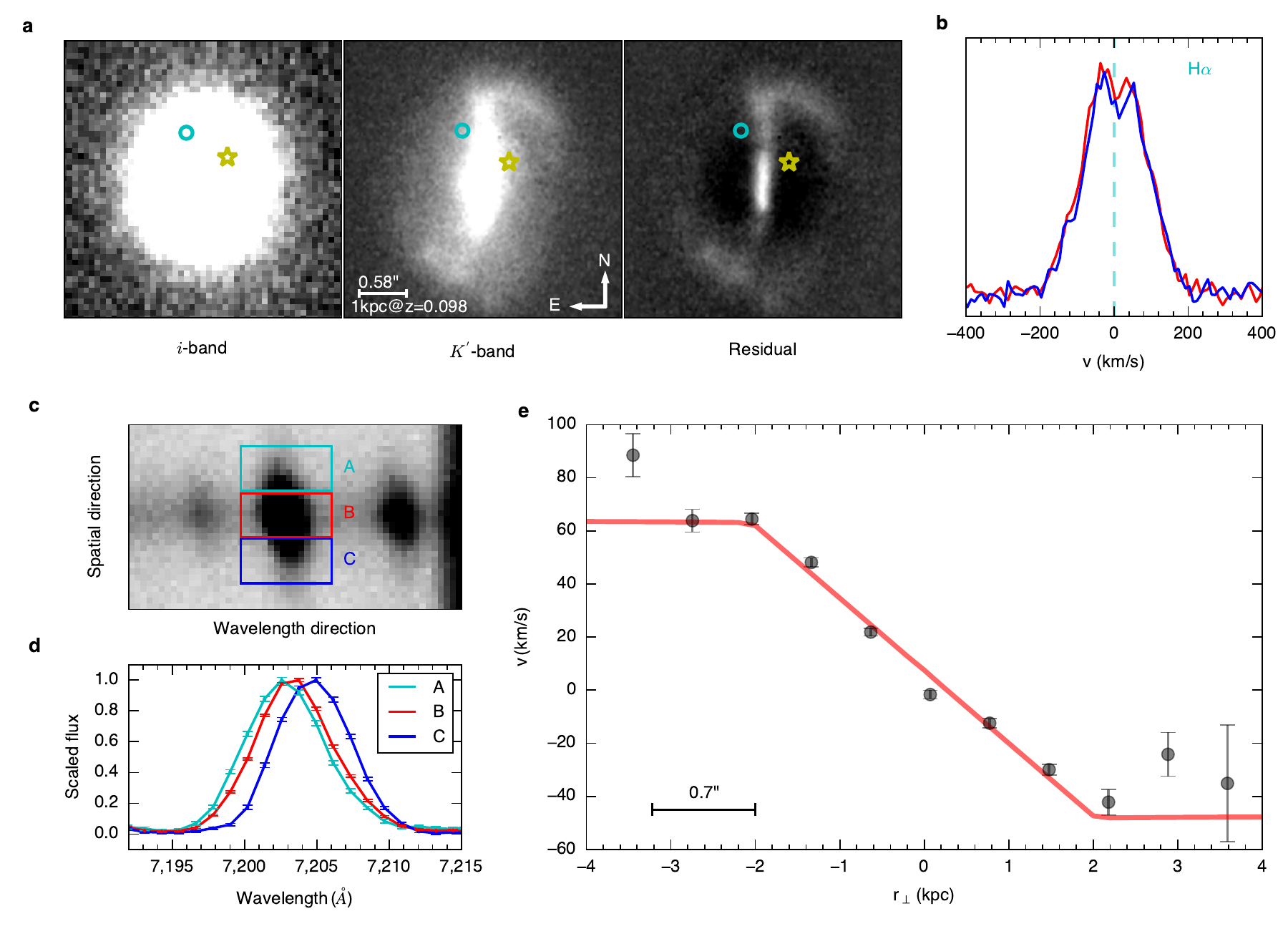}
\caption{\textbf{Properties of the host galaxy at $z=0.098$ in the optical and near-infrared.} {\bf (a)}  \REV{$i$-band and $K^{\prime}$-band FRB~20201124A host-galaxy images by LRIS and NIRC2, respectively, and the residual $K^{\prime}$-band image after subtracting the disc component.} 
The EVN localisation of FRB~20201124A is indicated with the cyan circle, which is in 60\,mas radius, i.e. 4 times the uncertainty. 
The centre of the background galaxy ($z=0.553$) is shown as the yellow asterisk. {\bf (b)} The H$\alpha$ double-peaked profile revealed in the medium-resolution ESI spectrum. Blue and red are for two different orders of the echelle spectrum. 
{\bf (c)} \REV{2D spectroscopic image by LRIS around the H$\alpha$ emission line}. 
A wavelength-dependent variation is clearly seen in the spatial direction. {\bf (d)} The H$\alpha$ lines extracted from three different regions, which correspond to the three rectangles \REV{in panel {\bf (c)}} of the galaxy along the slit. {\bf (e)} The velocities at different projected distances in the slit direction relative to the continuum centre. The red line is the best-fit result of a simple rotation model
The LRIS spectroscopic observations were taken with seeing of $0.7\arcsec$ (black bar), which sets the spatial resolution.
\label{fig:opticalobs01} }
\end{figure}

\begin{figure} 
\centering
\includegraphics[width=1.0\textwidth]{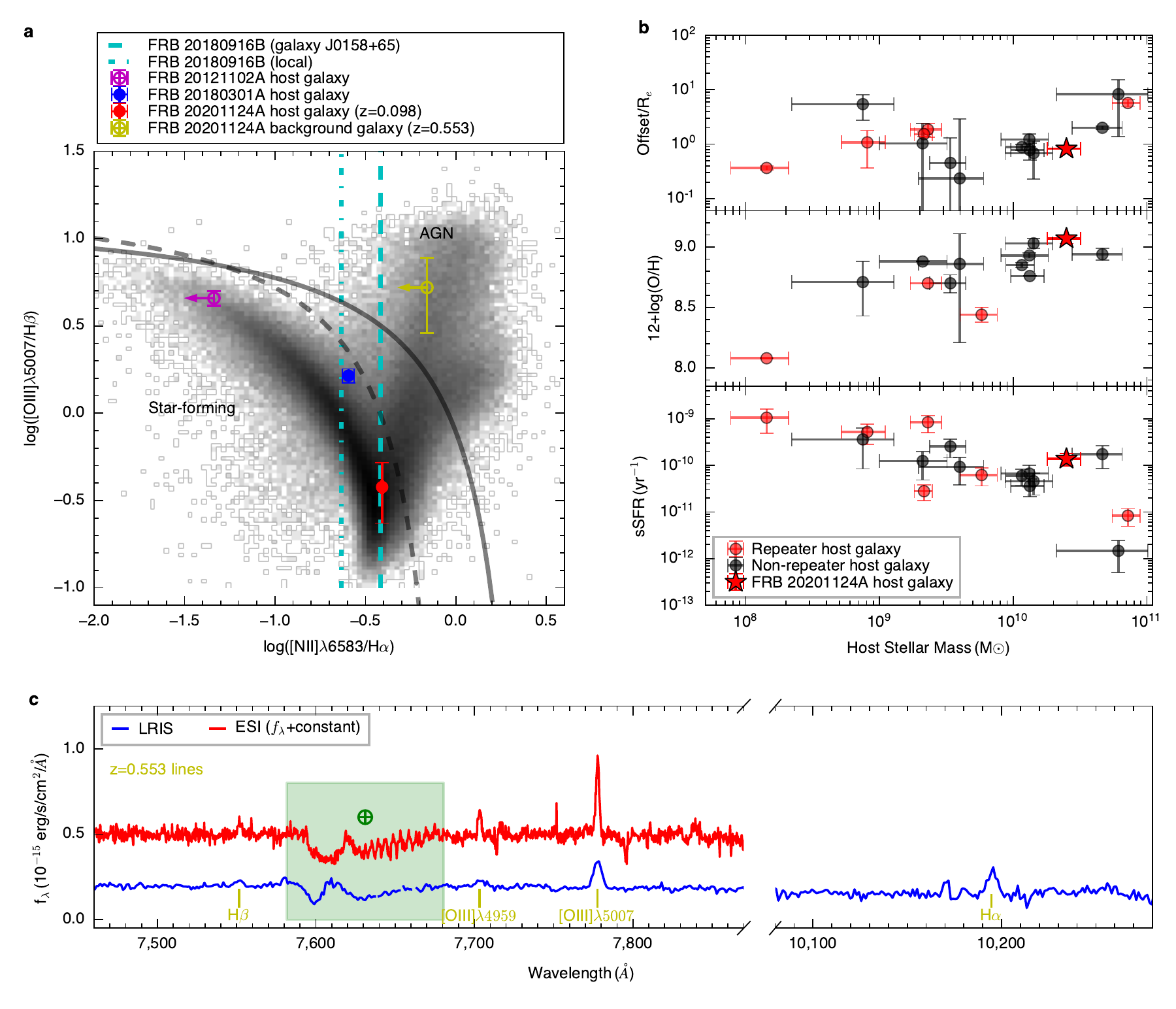}
\caption{\textbf{Properties of the galaxies and comparisons with other FRB hosts} 
{\bf (a)} FRB repeaters' hosts in the BPT diagram plotted with the SDSS DR8 MPA-JHU sample (black); parameter spaces of galaxies dominated by star formation and active galactic nuclei are separated by the black dashed and solid lines, respectively\cite{Kewley2001, Kauffmann2003}.
The host and background galaxies of FRB 2020124A are shown in red and yellow, respectively. {\bf (b)} The properties (FRB--galaxy offset in units of galaxy effective radius $R_e$, gas-phase metallicity, sSFR, and stellar mass) of the FRB~20201124A host galaxy (red star) compared with a literature sample of FRB hosts (available at \url{https://web.archive.org/web/20211015143528/https://frbhosts.org/\#explore}) shown with dots (black, non-repeaters; red, repeaters). {\bf (c)} Emission lines from the background galaxy at $z=0.553$ in the LRIS (blue) and ESI (red) spectra with regions contaminated by Earth's atmosphere marked in green. 
\label{fig:opticalobs02} }
\end{figure}

\clearpage

\clearpage
\bibliography{ms}

\end{document}